\documentclass[12pt,a4paper]{article}
\usepackage{amsmath}
\usepackage{amssymb}
\usepackage{amsthm}
\usepackage{float}
\usepackage{amsfonts}
\usepackage{subfig}
\usepackage{graphicx}
\usepackage{verbatim}
\usepackage[left=2cm,right=2cm,top=3cm,bottom=2.5cm]{geometry}
\usepackage[numbers]{natbib}
\usepackage{notoccite}
\usepackage[utf8]{inputenc}
\usepackage[usenames,dvipsnames,svgnames]{xcolor}
\usepackage[colorlinks=true,
      linkcolor=red,
      urlcolor=gray,
      citecolor=blue]{hyperref}

\def\myalign#1{%
  \def\trule{\noalign{\smallskip\hrule\medskip}}
  \def\nebc{\nearrow\bigcup}
  \def\sebc{\searrow\bigcup}
  \def\pminf{{}_{-\infty}|^{+\infty}}
  \let\Inf\infty
  \def\amp{&} 
  \vbox{\mathsurround0pt\openup1\jot
    \halign{%
      &$\displaystyle##\hfil\tabskip0pt$&\amp##\tabskip1em\crcr
      \noalign{\hrule height1pt\smallskip}#1\noalign{\smallskip\hrule height1pt}\crcr}}}

\def\ber {\begin{eqnarray}}
\def\eer {\end{eqnarray}}
\newcommand{\be}{\begin{equation}}
\newcommand{\ee}{\end{equation}}

\begin{document}

\begin{center}
 \textbf{Perturbations of quasi-Newtonian universes in scalar-tensor gravity}
\end{center}
\hfill\newline
Heba Sami$^{1}$  and Amare Abebe$^{1}$\\
Email: hebasami.abdulrahman@gmail.com\\
\hfill\newline
$^{1}$ Center for Space Research, North-West University, South Africa\\

\begin{center}
Abstract 
\end{center}
  In this contribution, we consider the equivalence between $f(R)$  gravity and scalar-tensor theories to study the evolution of scalar cosmological perturbations in the $1 + 3$ covariant formalism for the classes of shear-free cosmological dust models with irrotational fluid flows. The $f(R)$ gravity is considered to be a subclass of Brans-Dicke models, we gave an overview on the equivalence between $f(R)$ gravity and scalar-tensor theories.  We use the $1 + 3$ covariant formalism to present the covariant linearised evolution and constraint equations. We then derive the integrability conditions describing a consistent evolution of the linearised field equations of quasi-Newtonian universes in the scalar- tensor theories of gravity.  Finally, we derive the evolution equations for the density and velocity perturbations of the quasi-Newtonian universe. We  apply the  harmonic decomposition and we explore the behaviour of the matter density contrast  by considering $R^{n}$ models.  We introduce the so-called quasi-static
approximation to study the approximated solutions on small scales. The growth of the matter density contrast for both short- and long -wavelength modes has been examined by applying certain assumptions of the initial conditions.\\
\newline
$keywords:$ $f(R)$ gravity, scalar field, quasi-Newtonian cosmologies, perturbations.
\newline
$PACS:$ 04.50.Kd, 98.80.Jk, 98.80.-k, 95.36.+x, 98.80.Cq

\section{Introduction}
In the late 1990s, astronomical observations indicated that the Universe is expanding in an accelerating way and these observations entirely changed our way of thinking and it revived theories containing a cosmological constant. Dark energy closely resemble a cosmological constant, and there is a belief  that the dark energy  with negative pressure are responsible for this late-time acceleration however, the dark energy is probably the most extensively speculated candidate in recent years, but it is merely a phenomenological addendum whose existence has not been predicted from the Big Bang/ inflationary cosmology \cite{sotiriou2006f, de2010f}. Its dynamical nature is hardly understood, and none of the variant models put forward have been convincingly viable so far. The late-time accelerating phase comes to be added up to an early-time accelerated epoch which known as the inflationary epoch. This epoch is required to solve the flatness, horizon and the inhomogeneity problems and how the primordial fluctuations seeded the formation of structure on large-scales. These problems could not be explained by the $\Lambda$CDM model, consequently, there are exist several approaches to the theoretical description of these problems, one of the most common such approaches comes in the form of the introduction of new matter or scalar field contributions to the action of general relativity. The second comes in the form of modifications to the theory of gravity. The modified gravity theories provide the very natural gravitational alternative for dark energy and they are extremely attractive in the applications for late-time acceleration. There are different alternatives to GR, namely modified theories of gravity such as $f(R)$ theories of gravity, Dvali-Gabadadze-Porrati gravity \cite{dvali20004d}, Braneworld gravity, the Einstein-ether theory, $f(R)$ theories of gravity and scalar-tensor theory which are the best-known alternative to general relativity. Among these gravitational theories, in this paper we present the $f(R)$ theory as scalar-tensor theory of gravitation. $f(R)$ theories were first proposed by Buchdal \cite{buchdahl1970non} in 1970, They are fourth-order theories of gravity which considered to be one of the most studied alternative theories.They are based on the modification of the gravitational action and they are generally obtained by including higher-order curvature invariants in the Einstein-Hilbert action, or by making the action non-linear in the Ricci curvature $R$ and/or contain terms involving combinations of derivatives of $R$ \cite{gidelew2013beyond}. The scalar-tensor theory of gravity is one of the most accepted alternatives to Einstein's theory of general relativity \cite{fujii2003scalar}. It was invented first by Jordan \cite{jordan1955schwerkraft} in 1955 and then taken over by Brans and  Dicke \cite{brans1961mach} to try to explain the gravitational interaction through both a scalar field and a tensor field. It is a higher-order theory, where degrees of freedom to explain different scenarios are present. Some of those degrees of freedom are scalar field, the coupling constant and cosmological constant as well \cite{fujii2004some, ntahompagaze2017f}. The way the scalar field enters the arena of the scalar–tensor theory is not simple. It does so through what is known as a non-minimal coupling term. The Brans-Dicke theory \cite{brans1961mach} is one of the special classes of the scalar-tensor theory, where the coupling parameter $\omega (\phi)$ is considered to be constant, the coupling parameter $\omega (\phi)$ is supposed to be independent of the scalar field $\phi$. In this paper, we present $f(R)$ gravity theory as a subclass of scalar-tensor theory. We consider the covariant form of the field equations of $f(R)$ gravity as a subclass of scalar-tensor  to study the  linear cosmological perturbations. There are two approaches to study the cosmological perturbations namely, the metric based approach \cite{lifshitz1992gravitational, bardeen1980gauge, kodama1984cosmological, bertschinger2000cosmological, dunsby1991gauge, dunsby1992covariant, gidelew2013beyond, hwang1991large, mukhanov1992theory} and the $1+3$ covariant approach \cite{ hawking1966perturbations, ellis1989covariant}. In the $1+3$  covariant approach, the perturbations defined describes true physical degree of freedom, it is formulated at the level of variables that are covariantly defined in the real universe, and are exactly gauge-invariant by construction \cite{challinor2000microwave}. This approach has been used recently to study the cosmological perturbations for different contexts of modified gravity and GR \cite{ dunsby1991gauge, ellis1989covariant}. \\
\newline
This paper is organised as follows: in Sections \ref{Sec1} and \ref{Sec2} respectively, we review the $1+3$ covariant approach, Kinematics quantities and the field equations in $f(R)$ theory of gravity and we review $f(R)$ theory as a subclass of scalar-tensor theory. In Sections \ref{Sec3} and \ref{Sec4} respectively, we study the quasi-Newtonian models in scalar-tensor theory and we derive the integrability conditions that describes a consistent evolution of the linearised field equations of the quasi-Newtonian universes. In Section \ref{sec5}, we define the gradient variables that describe the cosmological perturbations and derive the linear evolution equations for matter and  scalar field perturbations and the harmonic decomposition approach is applied to these equations. In Section \ref{sec7}, we analyse the growth of the matter density contrast by considering $R^n$ models and by solving the whole system of perturbation equations numerically. We introduce the so-called quasi-static approximation to admit the approximated solutions on small scales. Section \ref{sec8} is devoted for discussions and conclusions.
\newpage
\section{ The $1+3$ covariant formalism}\label{Sec1}
In the $1+3$ covariant approach, space-time is split into space and time; it is divided into  perpendicular $4$-velocity field vector $u^{a}$ and  foliated hyper-surfaces. The vector field $u^{a}$ is given as \cite{ellis1989covariant, challinor2000microwave}\\ 

\be 
u^{a}= \dfrac{dx^{a}}{d\tau}\;, \hspace*{1cm} u^{a}u_{a}=-1\;,
\ee
where $\tau$ is proper time measured along the worldlines. In this approach the metric $g_{ab}$ is decomposed into the projected tensor $h_{ab}$ as follows:
\begin{equation}\label{1+3}
g_{ab}= h_{ab}- u_{a}u_{b}\;,\quad \mbox{with}\quad h^{a}_{c}h^{c}_{b}= h^{a}_{b}\;,\quad  h^{a}_{a}=3\;,\quad h_{ab}u^{b}=0\;.
\end{equation}
In this approach, the kinematic quantities which  describe all the kinematic features of the fluid flow can be obtained from irreducible parts of the decomposed $\nabla_{a}u_{b}$ as \cite{gidelew2013beyond, carloni2010conformal}
\begin{equation}\label{definitionofu}
\nabla_{a}u_{b}=\tilde{\nabla}_{a}u_{b}-u_{a}\dot{u}_{b}=\frac{1}{3}\theta h_{ab}+\sigma_{ab}+\omega_{ab}-u_{a}\dot{u}_{b}\; ,
\end{equation}
where the volume rate of expansion of the fluid is given as 
$\theta=3H(t)$
where $H(t)= \dot{a}/a$ is the Hubble which represents the rate of expansion at any time. 
The rate of distortion of the matter flow  is given as
$\sigma_{ab}=\tilde{\nabla}_{\langle a}u_{b\rangle}$
and it is the trace-free symmetric rate of the shear tensor $(\sigma_{ab}=\sigma_{(ab)},\sigma_{ab}u^{b}=0, \sigma^{a}_{a}=0)$. 
The relativistic acceleration vector is given as
$
\dot{u}^{a}=u^{b}\nabla_{b}u^{a}
$
and the skew-symmetric vorticity tensor that describes the rotation of the fluid relative to a non-rotating frame
$\omega_{ab}=\tilde{\nabla}_{[a}u_{b]}.$
The  energy-momentum tensor $T_{ab}$ is  decomposed using the $1+3$ covariant approach and it is given as \cite{gidelew2013beyond}
\begin{equation}\label{e}
T_{ab}=\rho u_{a}u_{b}+q_{a}u_{b}+u_{a}q_{b}+ph_{ab}+\pi_{ab}\; ,
\end{equation}
where $q_{a}$ and $\pi_{ab}$ are the heat flux and anisotropic pressure, respectively.
\section{Field equations}\label{Sec2}
The Einstein Field Equations (EFEs) can be derived from the Riemann geometry
(Ricci tensor, Ricci scalar and Einstein tensor) or from the Lagrangian in the Einstein-Hilbert action  and it is given as \cite{sotiriou2006f, sotiriou2010f}
\begin{eqnarray}\label{HA}
S_{EH}= \dfrac{1}{2}\int  \sqrt{-g}\Big(R+2\mathcal{L}_{m}(g_{\mu \nu},\psi)\Big)d^{4}x\; ,
\end{eqnarray}
where $\mathcal{L}_{m}$ is the matter Lagrangian, $g$ is the determinant of the metric tensor $g_{\mu \nu}$ and $\psi$ is the matter field. By replacing  the Ricci scalar $R$ with a function $f(R)$ in the Einstein-Hilbert action in Eq. \eqref{HA}, we will get the action of $f(R)$ gravity as \cite{joras2011some,gidelew2013beyond}
\begin{eqnarray}\label{FRA}
S_{f(R)}= \dfrac{1}{2\kappa}\int \sqrt{-g}\Big(f(R)+2\mathcal{L}_{m}(g_{\mu \nu},\psi)\Big)  d^{4}x\; .
\end{eqnarray}
One can apply different variational principles to the Einstein-Hilbert action Eq. \eqref{HA} in order to derive Einstein's field equations. They can be divided into three formalisms based on which variational principle is used, the metric, Palatini and metric-affine $f(R)$ formalisms. In the metric (or second-order) formalism, the action is varied with respect to the metric $g_{ab}$ \cite{de2010f, sotiriou2007modified, clifton2012modified, vitagliano2010dynamics, capozziello2011extended}. In the Palatini (or first-order) formalism, the metric and the connection are assumed to be independent variables and the action varies  with respect to both of them. In the metric-affine formalism, the metric and the connection are independent, as in
the case of the Palatini formalism. However, in this metric-affine theories of gravity there is a direct coupling between matter and connection \cite{hehl1978metric, vitagliano2011dynamics}. The action varies with respect to the metric and the affine connection neglecting the assumption that the matter action is independent of the connection. Here, we will be interested in metric-formalism.  The field equations coming from the action in Eq. \eqref{FRA} are given as \cite{de2010f, sotiriou2007modified, clifton2012modified, vitagliano2010dynamics, capozziello2011extended}
\begin{equation}\label{metricFE}
 T^{R}_{\mu \nu}= f^{'}(R)R_{\mu \nu} -\dfrac{1}{2}f(R)g_{\mu \nu}-\Big(\nabla_{\mu} \nabla_{\nu}-g_{\mu \nu} \nabla_{\mu}\nabla^{\mu}\Big)f^{'}(R)\; ,
\end{equation}
where $f=f(R)$ and  $f^{'}= \dfrac{df}{dR},$ and the energy-momentum tensor of matter is given as
$T^{m}_{\mu \nu}= -\dfrac{2\delta\left(\sqrt{-g}\mathcal{L}_{m}\right)}{\sqrt{-g} \delta g^{\mu \nu}}\;.$
From the last two terms of Eqs. \eqref{metricFE}, we notice that the field equations obtained in $f(R)$ are of fourth-order partial differential equations in the metric. However, the fourth-order terms vanish when $f^{'}$ is a constant, i.e. for an action which is linear in  $R$. Thus, it is straightforward for these equations to reduce to the Einstein equations once $f(R)= R$ \cite{sotiriou2007modified}.
\subsection{$f(R)$ theory as scalar-tensor theory of gravitation}
Scalar-tensor theories have been considered as the best-known alternative to GR. The general form of the action of the scalar-tensor theory is given as follows:
\begin{equation}\label{scalartensorAC}
S_{ST}=\dfrac{1}{2} \int d^{4}x \sqrt{-g} \Big[f(\phi)R- g(\phi)\nabla_{\mu} \phi \nabla^{\mu} \phi- 2\Lambda(\phi) + \mathcal{L}_{m}(g_{\mu \nu} \psi)\Big] \; ,
\end{equation}
where  $f$, $g$ and $\Lambda$ are some arbitrary functions of the scalar field $\phi$  \cite{clifton2012modified, bergmann1968comments, nordtvedt1970post, wagoner1970scalar}. By a redefinition of the scalar field $\phi$ we can now set $f(\phi)\rightarrow \phi$, without loss of generality. The Lagrangian density in Eq. \eqref{scalartensorAC} can then be written as \cite{ clifton2012modified}
 \begin{equation}\label{scalartensorAC2}
S_{ST}= \dfrac{1}{2}\int d^{4}x \sqrt{-g} \Big[\phi R- \dfrac{\omega(\phi)}{\phi}\nabla_{\mu} \phi \nabla^{\mu} \phi- 2\Lambda(\phi) + \mathcal{L}_{m}(g_{\mu \nu} \psi)\Big]\; , 
\end{equation}
where $\omega(\phi)$ is an arbitrary function, often referred to as the coupling parameter. The variation of the action in Eq. \eqref{scalartensorAC2} with respect to the metric $g_{\mu \nu}$ gives the field equations as
\begin{equation}\label{scalartensorFE}
\phi G_{\mu \nu}+ \Big[\square\phi +\dfrac{1}{2}\dfrac{\omega}{\phi}(\nabla \phi)^{2}+ \Lambda \Big]g_{\mu \nu}- \nabla_{\mu} \nabla _{\nu}\phi -\dfrac{\omega}{\phi}\nabla_{\mu}\phi \nabla_{\nu}\phi = T_{\mu \nu}\; .
\end{equation}
Now, as well as the metric tensor $g_{\mu \nu}$, these theories also contain the dynamical scalar field $\phi$, and so we must vary the action derived from Eq. \eqref{scalartensorAC2} with respect to this additional degree of freedom. After eliminating $R$ with the trace of Eq. \eqref{scalartensorFE}, this yields \cite{clifton2012modified}
\begin{equation}\label{scalartensorFE2}
(2\omega+3)\square\phi+ \omega^{'}(\nabla \phi)^{2}+ 4\Lambda-2\phi \Lambda^{'}= T\; ,
\end{equation}
where primes here denote differentiation with respect to $\phi$. These are the field equations of the scalar-tensor theories of gravity. The Brans-Dicke theory \cite{brans1961mach} is one of the special classes of the scalar-tensor theory, by setting $\omega(\phi)$ to be a constant and dropping the cosmological constant in the action given in Eq. \eqref{scalartensorAC2}, we write the BD action as
\begin{equation}\label{BDAct}
S_{BD}=\frac{1}{2}\int d^{4}x \sqrt{-g}\left[\phi R-\frac{\omega}{\phi}\nabla_{\mu}\phi\nabla^{\mu}\phi+2\mathcal{L}_{m}\right]\; .
\end{equation}  
The Brans-Dicke theory  is based on certain assumptions. The matter Lagrangian $\mathcal{L}_{m}$ is assumed not to contain the scalar field $\phi$ and the scalar field is also assumed to be cosmic time dependent only \cite{fujii2003scalar}.\\
When a homogeneous and isotropic universe is assumed, the Einstein-Hilbert fundamental constant of gravitational $G$ is related to  the gravitation constant $G_1$ in the BD theory by \cite{narlikar2002introduction}
\begin{equation}\label{BDEH}
G_{1}= \dfrac{4+2\omega}{3+2\omega} G\;. 
\end{equation}
When the coupling constant $\omega$ tends to infinity in Eq. \eqref{BDEH}, the BD theory is reduced to Einstein gravity \cite{narlikar2002introduction}.  A clear illustration of how $f(R)$
theories are classified as a subclass of a scalar-tensor theory is in the Brans-Dicke theory for the case of the coupling constant $\omega =0$ in Eq. \eqref{BDAct}. $f(R)$ theory is related to scalar-tensor theory by defining the scalar field $\phi$ to be a function of Ricci scalar $R$, by which each $f(R)$ has a potential $U(\phi)$. 
 This work makes use of the equivalence between $f(R)$ theory in its metric formalism  and the scalar-tensor theories of gravity.
In this equivalence the scalar-tensor action takes the form \cite{gidelew2013beyond,frolov2008singularity,sami2017reconstructing}
\begin{equation}\label{scalartensorAC3}
S_{f(\phi)}=  \dfrac{1}{2\kappa}\int \sqrt{-g}\Big(f(\phi(R))+2\mathcal{L}_{m}(g_{\mu \nu},\psi)\Big) d^{4}x\; ,
\end{equation}
where $f(\phi(R))$ is  now a function of $\phi(R)$ and we define the scalar field $\phi$ to be 
\begin{equation} \label{equivalence}
\phi = f^{'}-1\; .
\end{equation}
Here the prime indicates differentiation with respect to $R$ and the scalar field $\phi$ should be invertible \cite{sotiriou2010f, clifton2012modified, faulkner2007constraining}. This definition allows $\phi$ to be an extra degree of freedom that vanishes in GR. Therefore, by comparing the scalar-tensor action in Eq. \eqref{scalartensorAC3} to that action of the BD in Equation \eqref{BDAct} for the case of vanishing coupling constant $\omega=0$, we see that the action of the BD results in generalized field equations that describe the same cosmological dynamics as the action of the scalar-tensor in Eq. \eqref{scalartensorAC3}. These two actions are dynamically  equivalent, thus  $f(R)$ theory is a special case of the scalar-tensor theory.
The field equations of the action in Eq. \eqref{equivalence} can be now written as \cite{gidelew2013beyond}
\begin{equation}\label{equivalenceFE}
G_{ab} = \dfrac{T^{m}_{ab}}{(\phi+1)}+ \dfrac{1}{(\phi+1)}\left( \dfrac{1}{2}g_{ab}\left( f-(\phi+1)R\right)+ \nabla_{a}\nabla_{b} \phi-g_{ab}\nabla_{c}\nabla^{c}\phi\right)\; .
\end{equation}
By decomposing the right hand side of Eq. \eqref{equivalenceFE} into two cosmological fluids (the standard matter and scalar field), Eq. \eqref{equivalenceFE} can be rewritten as 
\begin{equation}\label{equivalenceFE2}
G_{ab}= \tilde{T}^{m}_{ab}+ T^{\phi}_{ab}\; ,
\end{equation}
where $\tilde{T}^{m}_{ab}= \dfrac{T^{m}_{ab}}{(\phi+1)}$ is the energy-momentum tensor for the effective matter fluid and 
\begin{equation}\label{EMSCALAR}
T^{\phi}_{ab}= \dfrac{1}{(\phi+1)}\left[ \dfrac{1}{2}g_{ab}\left( f-(\phi+1)R\right)+ \nabla_{a}\nabla_{b} \phi-g_{ab}\nabla_{c}\nabla^{c}\phi\right]\; ,
\end{equation}
is the energy momentum tensor for the scalar field.
\subsection{Background thermodynamics of scalar-tensor gravity}
The linearised thermodynamic quantities for the scalar field are the
energy density $\rho_{\phi}$, the  pressure $p_{\phi}$, the energy flux $q^{\phi}_{a}$ and the anisotropic pressure $\pi^{\phi}_{ab}$, respectively given by
\begin{eqnarray}\label{8}
&&\rho_{\phi}= \frac{1}{(\phi +1)} \Big[\frac{1}{2}\Big( R(\phi +1)-f\Big)- \theta \dot{\phi} +  \tilde{\nabla}^{2}\phi \Big]\;,\\
&&\label{9} p_{\phi }= \frac{1}{(\phi +1)}\Big[ \frac{1}{2} \Big( f- R(\phi+1)\Big) + \ddot{\phi} - \frac{\dot{\phi} \phi^{\prime \cdot}}{\phi^{'}} + \frac{\phi^{''} \dot{\phi}^{2}}{\phi^{'2}}+ \frac{2}{3}(\theta \dot{\phi} - \tilde{\nabla}^{2} \phi) \Big]\;,\\ 
\label{10}
&& q^{\phi}_{a}= -\frac{1}{(\phi+1)}\Big[ \frac{\dot{\phi}^{'}}{\phi^{'}} - \frac{1}{3} \theta \Big]\tilde{\nabla}_{a} \phi\;,\\
&&\label{101}\pi ^{\phi}_{ab} = \frac{\phi^{'}}{(\phi+1)} \Big[\tilde{\nabla}_{\langle a} \tilde{\nabla}_{b \rangle}R - \sigma_{ab} \Big(\frac{\dot{\phi}}{\phi^{'}}\Big)\Big]\;.
\end{eqnarray}
The total ({\it effective})  energy density, isotropic pressure, anisotropic pressure and heat flux of standard matter and scalar field combination are given by
\begin{eqnarray}\nonumber
\rho \equiv\frac{\rho_{m}}{(\phi+1)}+ \rho_{\phi}, \hspace{.4cm} p \equiv \frac{p_{m}}{(\phi+1)}+ p_{\phi} \;, \hspace{.4cm} \pi_{ab}\equiv \frac{\pi^{m}_{ab}}{(\phi+1)}+ \pi^{\phi}_{ab}, \hspace{.4cm} q_{a}\equiv \frac{q^{m}_{a}}{(\phi+1)}+ q^{\phi}_{a}\;.
\end{eqnarray}
The Freiedmann equations are presented as follow
\begin{equation}\label{e1}
H^{2}= \dfrac{1}{3(\phi+1)} \Big(\rho_{m}+ \dfrac{1}{2}(R(\phi+1)-f)+ \theta \dot{\phi}+\nabla^{2}\phi \Big)\;, 
\end{equation}
\begin{equation}\label{e2}
2\dot{H}+3H^{2}= \dfrac{-1}{(\phi+1)} \Big( \dfrac{1}{2} (f-R(\phi+1))+\ddot{\phi}- \dfrac{\dot{\phi}\phi^{\prime \cdot}}{\phi’}+ \dfrac{\phi'' \dot{\phi}^{2}}{\phi'^2}+\dfrac{2}{3}\theta\dot{\phi}-\dfrac{2}{3}\nabla^{2}\phi\Big)\;,
\end{equation}
therefore, the Friedmann Eq. \eqref{e1} can be written as
\begin{eqnarray} \label{e3}
 1= \tilde{\Omega}_{m}+ \mathcal{X}+\mathcal{Y}+ \dfrac{\tilde{\nabla}^{2}\phi}{3H^2 (\phi+1)} \;. 
\end{eqnarray}
For simplicity, we have introduced the following dimensionless variables 
\begin{eqnarray}\label{e4}
 \tilde{\Omega}_{m}= \dfrac{\rho_{m}}{3H^{2}(\phi+1)},  \hspace*{.3cm}  \mathcal{X}= \dfrac{1}{6H^2 (\phi+1)}(R(\phi+1)-f), \hspace*{.3cm} \mathcal{Y}= \dfrac{\dot{\phi}}{H(\phi+1)}\;.
\end{eqnarray}
Where $\tilde{\Omega}_{m}$ is the fractional energy density of effective matter like fluid. In the FLRW spacetime universe, the  Friedmann and Raychaudhuri  equations that govern the expansion history of the Universe can be written as follows \cite{carloni2008evolution}:
\begin{eqnarray}\label{6}
&&\dot{\theta}+\frac{1}{3} \theta^{2}= -\dfrac{1}{2}(\rho+3p)+\nabla^{a}A_{a}\;,\\&&\label{7}
\theta^{2}= \frac{3}{(\phi+1)}\Big[ \rho_{m} + \frac{R(\phi +1) - f}{2} + \theta \dot{\phi}\Big]\;,
\end{eqnarray}
\subsection{Covariant equations}\label{sec1}
Given a choice of $4$-velocity field $u^{a}$, the Ehlers-Ellis approach \cite{ellis1989covariant, ehlers1993contributions} employs only fully covariant quantities and equations with transparent physical and geometric meaning \citep{maartens98}. In such a treatment for any scalar quantity $X$ in the background we have
$$\tilde{\nabla}_{a}X=0\; ,$$
thus, by virtue of the Stewart-Walker Lemma \citep{stewart1974perturbations}, any quantity is considered to be gauge-invariant if it vanishes in the background. Therefore, the FLRW background is characterised by the following dynamics, kinematics and gravito-electromanetics  \citep{abebe2016integrability}:
\begin{eqnarray}\label{110}
&&\tilde{\nabla}_{a}\rho=0=\tilde{\nabla}_{a}p=\tilde{\nabla}_{a}{\theta}\;,\quad q_{a}=0=A_{a}=\omega_{a}\;\nonumber\\
&& \pi_{ab}=0= \sigma_{ab}=E_{ab}= 0=H_{ab}\;,\nonumber\\
\end{eqnarray}
where $E_{ab}$ and $H_{ab}$ are the gravito-electromagnetic fields responsible for tidal forces and gravitational waves. They are the ``gravito-electric'' and ``gravito-magnetic'' components of the Weyl tensor $C_{abcd}$  defined from the Riemann tensor $R^a_{bcd} $ as 
\begin{eqnarray}\label{weyl}
&&C^{ab}{}_{cd}=R^{ab}{}_{cd}-2g^{[a}{}_{[c}R^{b]}{}_{d]}+\frac{R}{3}g^{[a}{}_{[c}g^{b]}{}_{d]}\;,\\
&&\label{geweyl}
E_{ab}\equiv C_{agbh}u^{g}u^{h},~~~~~~~H_{ab}\equiv\dfrac{1}{2}\eta_{ae}{}^{gh}C_{ghbd}u^{e}u^{d}\;.
\end{eqnarray}
The covariant linearised evolution equations in the general case are given by \citep{maartens98, abebe2016integrability, maartens1997density} 
\begin{eqnarray}\label{4000}
&&\dot{{\theta}}= -\dfrac{1}{3}{\theta}^{2} - \dfrac{1}{2}(\rho+ 3p)+ \tilde{\nabla}_{a}A^{a}\;,  \hspace*{.4cm} \label{5}
\dot{\rho}_{m}= -\rho_{m}{\theta} -\tilde{\nabla}^{a}q^{m}_{a}\; ,\\
&&\label{6000}
\dot{q}^{m}_{a}= -\dfrac{4}{3}{\theta} q^{m}_{a}- \rho_{m} A_{a}\; ,  \hspace*{.4cm} 
\label{10}
\dot{\sigma}_{ab}=-\dfrac{2}{3}{\theta} \sigma_{ab}-E_{ab} + \dfrac{1}{2}\pi_{ab}+ \tilde{\nabla}_{\langle a}A_{b\rangle}\; ,\\
&&\label{9}
\dot{\omega}^{\langle a\rangle}= -\dfrac{2}{3}{\theta} \omega^{a}- \dfrac{1}{2}\eta^{abc}\tilde{\nabla}_{b}A_{c}\; ,\\
&&\label{7}
\dot{E}^{\langle ab\rangle}= \eta^{cd\langle a}\tilde{\nabla}_{c} H^{\rangle b}_{d}- {\theta} E^{ab} - \dfrac{1}{2}\dot{\pi}^{ab}-\dfrac{1}{2}\tilde{\nabla}^{\langle a}q^{b\rangle}- \dfrac{1}{6}{\theta} \pi^{ab}\; ,\\
&&\label{8}
\dot{H}^{\langle ab\rangle}= -{\theta} H^{ab}- \eta^{cd\langle a}\tilde{\nabla}_{c}E^{\rangle b}_{d}+ \dfrac{1}{2} \eta^{cd\langle a}\tilde{\nabla}_{c}\pi^{\rangle b}_{d}\;.
\end{eqnarray}
These evolution equations propagate consistent initial data on some initial $(t = t_{0})$
hypersurface $S_{0}$ uniquely along the reference time-like  consistency \citep{abebe2015irrotational}. They are constrained by the following linearised equations \citep{maartens98, abebe2016integrability, maartens1997density}:
\begin{eqnarray}\label{11}
&&C^{ab}_{0}\equiv E^{ab}- \tilde{\nabla}^{\langle a}A^{b \rangle}- \dfrac{1}{2} \pi^{ab}=0\; , \hspace*{.4cm} 
\label{12}
C^{a}_{1}\equiv \tilde{\nabla}_{b}\sigma^{ab}- \eta^{abc}\tilde{\nabla}_{b}\omega_{c}- \dfrac{2}{3}\tilde{\nabla}^{a}{\theta}+ q^{a}=0\; ,\\
&&\label{13}
C_{2}\equiv \tilde{\nabla}^{a}\omega_{a}=0\; , \hspace*{.4cm} 
\label{14}
C^{ab}_{3}\equiv\eta_{cd(}\tilde{\nabla}^{c}\sigma^{d}_{b)}+\tilde{\nabla}^{\langle a}\omega^{b\rangle}-H^{ab}=0\; ,\\
&&\label{15}
C^{a}_{5}\equiv\tilde{\nabla}_{b}E^{ab}+\dfrac{1}{2}\tilde{\nabla}_{b}\pi^{ab}-\dfrac{1}{3}\tilde{\nabla}^{a}\rho+\dfrac{1}{3}{\theta} q^{a}=0\;,\\&&
\label{16}
C^{a}_{b}\equiv\tilde{\nabla}_{b}H^{ab}+(\rho +p)\omega^{a}+\dfrac{1}{2}\eta^{abc}\tilde{\nabla}_{b}q_{a}=0\; .
\end{eqnarray}
These constraints restrict the initial data to be specified  and they must remain satisfied on any hypersurface $S_{t}$ for all comoving time $t$
\section{Quasi-Newtonian spacetimes}\label{Sec3}
 The importance of investigating the Newtonian limit for general relativity on cosmological contexts is that, there is a viewpoint that cosmology is essentially a Newtonian affair, with the relativistic theory only needed for examination of some observational relations. Most of the astrophysical calculations on the formation of large-scale structure in the universe rely on such a limit \cite{van1998quasi}. In \cite{van1998quasi}, a covariant approach to cold matter universes in quasi-Newton has been developed and it has been applied and extended in \cite{maartens1998newtonian} in order to derive and solve the equations governing density and velocity perturbations. This approach revealed the existence of integrability conditions in GR.\\
If a comoving  $4$-velocity $\tilde{u}^{a}$ is chosen such that, in the linearised form 
\begin{equation}\label{29}
\tilde{u}^{a}= u^{a}+v^{a}, \hspace{.3cm} v_{a}u^{a}=0, \hspace{.3cm} v_{a}v^{a}<<1\;,
\end{equation}
the  dynamics, kinematics and gravito-electromagnetics quantities Eq.  \eqref{110} undergo transformation. Here $v^{a}$ is the relative velocity  of the comoving frame with respect to the  observers in the quasi-Newtonian frame, defined such that it vanishes in the background. In other words, it is a non-relativistic peculiar velocity.  Quasi-Newtonian cosmological models are irrotational, shear-free dust spacetimes characterised by \cite{maartens98, abebe2016integrability}:
\begin{equation}\label{30}
 p_{m}=0\;, \hspace*{.3cm} q^{m}_{a}=\rho_{m} v_{a}\;, \hspace*{.3cm} \pi^{m}_{ab}=0\;,\omega_{a}=0\;, \hspace*{.3cm}  \sigma_{ab}=0\;.
\end{equation}
The gravito-magnetic constraint Eq. \eqref{14} and the shear-free and irrotational condition \eqref{30} show that the gravito-magnetic component of the Weyl tensor automatically vanishes:
\begin{equation}\label{32}
H^{ab}=0\;.
\end{equation}
The vanishing of this quantity implies  no  gravitational radiation in quasi-Newtonian cosmologies, and Eq. \eqref{16} together with Eq. \eqref{30} show that $q^{m}_{a}$ is irrotational and thus so is $v_{a}$:
\begin{equation}
\eta^{abc}\tilde{\nabla}_{b}q_{a}=0 = \eta^{abc}\tilde{\nabla}_{b}v_{a}\;.
\end{equation}
Since the vorticity vanishes, there exists a velocity potential such that
\begin{equation}\label{33}
v_{a}= \tilde{\nabla}_{a}\Phi\;.
\end{equation}
\section{Integrability conditions}\label{Sec4}
A constraint equation $C^{A}=0$ is said to evolve consistently with the evolution equations\cite{maartens98} if
\begin{equation}
\dot{C}^{A}= F^{A}_{B}C^{B}+G^{A}B_{a}\tilde{\nabla}^{a}C^{B}\; ,
\end{equation}
where $F$ and $G$ are quantities that depend on the kinematics, dynamics and gravito-electromagnetics quantities but not their derivatives. 
It has been shown \cite{van1997integrability} that the non-linear models are generally inconsistent if the silent constraint Eq. \eqref{32} is imposed, but that the linear models are consistent \cite{maartens98, abebe2016integrability}. Thus, a simple approach to the integrability conditions for quasi-Newtonian cosmologies follows from showing that these models are in fact a sub-class of the linearised silent models. This can  happen by using the transformation between the quasi-Newtonian and comoving frames.\\
The transformed linearised kinematics, dynamics and gravito-electromagnetic quantities from the quasi-Newtonian frame to the comoving frame are given as follows:
\begin{eqnarray}\label{34}
&&\tilde{\theta} =\theta +\tilde{\nabla}^{a}v_{a}\;,\hspace{.3cm}
\tilde{A}_{a}= A_{a}+ \dot{v}_{a}+\frac{1}{3}\theta v_{a}\;,\\&&\label{36}
\tilde{\omega}_{a}= \omega_{a}- \frac{1}{2}\eta_{abc}\tilde{\nabla}^{b}v^{c}\;,\hspace{.3cm}
\tilde{\sigma}_{ab}= \sigma_{ab}+ \tilde{\nabla}_{\langle a}v_{b\rangle}\;,\\&&
\label{38}
\tilde{\rho}= \rho, \hspace{.3cm} \tilde{p}=p, \hspace{.3cm} \tilde{\pi}_{ab}= \pi_{ab}, \hspace{.3cm} \tilde{q}^{\phi}_{a}= q^{\phi}_{a}\;, \hspace{.3cm}
\label{39}
\tilde{q}^{m}_{a}= q^{m}_{a}-(\rho_{m} +p_{m})v_{a}\;,\\&&\label{40}
\tilde{E}_{ab}= E_{ab}, \hspace{.3cm} \tilde{H}_{ab}= H_{ab}\;.
\end{eqnarray}
It follows from the above transformation equations that
\begin{eqnarray}\label{41}
&&\tilde{p}_m=0\;,\hspace*{.3cm} \tilde{q}^m_{a}=0=\tilde{A}_{a}=\tilde{\omega}_{a}\;, \\ \nonumber && \tilde{\pi}^m_{ab}=0=\tilde{H}_{ab}\;,\hspace*{.3cm}
\tilde{\sigma}_{ab}=\tilde{\nabla}_{\langle a}v_{b\rangle}\;,\hspace*{.3cm}\tilde{E}_{ab}= E_{ab}\;.
\end{eqnarray}
These equations describe the linearised silent universe except that the restriction on the shear in Eq. \eqref{41} results in the integrability conditions for the quasi-Newtonian models. Due to the vanishing of the shear in the quasi-Newtonian frame, Eq. \eqref{10} is turned into a new constraint 
\begin{equation}\label{44}
E_{ab}- \frac{1}{2}\pi^{\phi}_{ab}-\tilde{\nabla}_{\langle a}A_{b\rangle}=0\;.
\end{equation}
This can be simplified by using Eq. \eqref{9} and the identity for any scalar $\varphi$: 
\begin{equation}\label{45}
\eta^{abc} \tilde{\nabla}_{a}A_{c}=0 \Rightarrow A_{a}= \tilde{\nabla}_{a}\varphi\;.
\end{equation}
In this case $\varphi$ is the covariant relativistic generalisation of the Newtonian potential.
\subsection{First integrability condition}
Since Eq. \eqref{44} is a new constraint, we need to ensure its consistent propagation at all epochs and in all spatial hypersurfaces. Differentiating it with respect to cosmic time $t$ and by using Eqs. \eqref{101}, \eqref{10} and \eqref{16}, one obtains
\begin{eqnarray}\label{46}
\tilde{\nabla}_{\langle a}\tilde{\nabla}_{b\rangle}\Big( \dot{\varphi} +\frac{1}{3} \theta +\frac{\dot{\phi}}{(\phi+1)}\Big)+\Big(\dot{\varphi}+\frac{1}{3} \theta  +\frac{\dot{\phi}}{(\phi+1)}\Big)\tilde{\nabla}_{a}\tilde{\nabla}_{b}\varphi =0\;,
\end{eqnarray}
which is the first integrability condition for quasi-Newtonian cosmologies in scalar-tensor theory of gravitation and it is a generalisation of the one obtained in \cite{maartens98}, {\it i.e.}, Eq.\eqref{46} reduces to an identity for the generalized van Elst-Ellis condition  \cite{maartens98, abebe2016integrability, van1998quasi}
\begin{equation}\label{47}
\dot{\varphi}+\frac{1}{3}\theta= -\frac{\dot{\phi}}{(\phi+1)}\;.
\end{equation}
Using Eq. \eqref{4000} with the time evolution of the modified van Elst-Ellis condition, we obtain the covariant modified Poisson equation in scalar-tensor gravity as follows:
\begin{equation}\label{49}
\tilde{\nabla}^{2}\varphi= - (3\ddot{\varphi}+\theta\dot{\varphi})+\dfrac{1}{2(\phi+1)}\Big( \rho_{m} -\big(R(\phi+1)-f\big)-\theta \dot{\phi} +9\ddot{\phi} - \frac{3 \dot{\phi} \phi^{\prime \cdot}}{\phi’} + \frac{3 \phi^{\prime \prime}\dot{\phi}^{2}}{ \phi’^2}- \frac{6 \dot{\phi}^{2}}{ (\phi+1)}- \tilde{\nabla}^{2}\phi\Big)\;.
\end{equation}
The evolution equation of the $4$-acceleration $A_{a}$ can be shown, using Eqs. \eqref{47} and \eqref{11} to be
\begin{eqnarray}\label{49}
\dot{A}_{a}+\Big(\frac{2}{3}\theta+\frac{\dot{\phi}}{(1+\phi)}\Big)A_{a}+\frac{1}{2(1+\phi)}\Big( \rho_{m} v_{a}+ \Big(\frac{1}{3}\theta- \frac{\dot{\phi}^{\prime}}{\phi^{\prime}}- \frac{2\dot{\phi}}{(1+\phi)}\Big)\tilde{\nabla}_{a}\phi\Big)+ \dfrac{1}{(\phi+1)}\tilde{\nabla}_{a} \dot{\phi}=0\;.
\end{eqnarray}
\subsection{Second integrability condition}
There is a second integrability condition arising by checking for the consistency of the constraint \eqref{44} on any spatial hyper-surface of constant time $t$. By taking the divergence of \eqref{44} and by using the identity \eqref{a4} which holds for any projected vector $A_{a}$, and by using Eq. \eqref{45}  it follows that:
\begin{eqnarray}\label{52}
&&\tilde{\nabla}_{a}\rho_{m}- \Big( \dot{\phi}+\frac{2}{3} (\phi+1)\theta\Big)\tilde{\nabla}_{a}\theta+ \frac{1}{(\phi+1)}\Big( \frac{f}{2}-\rho_{m}+\theta\dot{\phi}  - \tilde{\nabla}^{2}\Big)\tilde{\nabla}_{a}\phi \nonumber \\&&-2(\phi+1)\tilde{\nabla}^{2}(\tilde{\nabla}_{a}\varphi) -2\Big(\rho_{m}+ \frac{1}{2}\big(R(\phi+1) f \big)-\theta\dot{\phi}-\frac{1}{3} \theta^{2}(\phi+1)\Big)\tilde{\nabla}_{a}\varphi \nonumber\\&&-3\phi^{\prime} \tilde{\nabla}_{a} \big( \tilde{\nabla}_{b} \tilde{\nabla}_{c}\big)R =0\;,
\end{eqnarray}
which is the second integrability condition and in general it appears to be independent of the first integrability condition \eqref{46}.
By taking the gradient of Eq. \eqref{47} and using Eq. \eqref{11}, one can obtain the peculiar velocity:
\begin{equation}\label{53}
v_{a}= -\frac{1}{\rho_{m}}\Big[ 2(\phi+1)\tilde{\nabla}_{a}\dot{\varphi}+ \Big( \frac{\dot{\phi}^{\prime}}{\phi^{\prime}}-\dfrac{1}{3}\theta^{2}+\frac{1}{(\phi+1)}\Big)\tilde{\nabla}_{a}\phi +2\tilde{\nabla}_{a}\dot{\phi}\Big]\;.
\end{equation} 
which evolves according to
\begin{equation}\label{54}
\dot{v_{a}}+\frac{1}{3}\theta v_{a}= -A_{a}\;.
\end{equation}
\section{Cosmological perturbations}\label{sec5}
In the previous section, we showed how imposing special restrictions to the linearized perturbations of FLRW universes in the quasi-Newtonian setting  result in the integrability conditions. These integrability conditions imply velocity and acceleration propagation equations resulting from the  generalised van Elst-Ellis condition for the acceleration potential in scalar-tensor theories. In this section, we show how one can obtain the velocity and density perturbations via these propagation equations, thus generalizing GR results obtained in \cite{maartens98}.\\
\newline
We define the variables that characterise scalar inhomogeneities the matter
energy density, expansion, peculiar velocity, acceleration as well as the scalar fluid and scalar field momentum, respectively, as follows:
\begin{eqnarray}\label{Delta}
&&\Delta^{m}= \frac{a^{2} \tilde{\nabla}^{2}\rho_{m}}{\rho_{m}}\; , \hspace{.5cm}
\label{Z}
Z= a^{2}\tilde{\nabla}^{2}\theta \; , \hspace{.5cm}
\label{V}
V^{m}=a^{2}\tilde{\nabla}^{a}v_{a}\; , \nonumber \\&&
\label{Ascalar}
\mathcal{A}= a^{2}\tilde{\nabla}^{a}A_{a}\; , \hspace{.5cm}
\label{phiscalar}
\Phi= a^{2}\tilde{\nabla}^{2}\phi\; \hspace{.5cm}
\label{psiscalar}
\Psi= a^{2}\tilde{\nabla}^{2}\dot{\phi}\; .
\end{eqnarray}
\subsection{First- and second- order evolution equations}
Due to the above  definitions of the scalar gradient variables, here we present  the first- and  the second-order evolution equations to demonstrate the growth of perturbations. The system of equations governing the evolutions of these scalar fluctuations are given as follows
\begin{eqnarray} \label{dotz}
&&\dot{Z}+\Big(\dfrac{2}{3}\theta + \dfrac{\dot{\phi}}{2(\phi+1)}\Big)Z+ \dfrac{\rho_{m}}{2(\phi+1)}\Delta^{m}+ \dfrac{\theta}{2(\phi+1)} \dot{\Phi} +\dfrac{1}{2(\phi+1)} \nabla^{2}\Phi \nonumber\\ && -\dfrac{1}{2(\phi+1)^{2}}\Big\lbrace2 \rho_{m} + (Rf'-f)-2\theta \dot{\phi}+ 3(\phi+1)\Big(\frac{\ddot{\phi} \phi^{\prime \cdot}}{ \phi^{\prime} \dot{\phi}} - \frac{ \dddot{\phi}}{ \dot{\phi}} + \frac{ \phi^{\prime \cdot \cdot}}{ \phi^{\prime}} -\frac{ \phi^{\prime \cdot 2}}{\phi^{\prime 2}}\nonumber \\&& - \frac{\phi ^{\prime \prime \cdot} \dot{\phi}}{\phi^{\prime2} }- \frac{2 \phi^{\prime \prime} \ddot{\phi}}{ \phi^{\prime 2}} + \frac{2 \phi^{\prime \prime} \phi^{\prime \cdot} \dot{\phi}}{ \phi^{\prime3}}\Big) \Big\rbrace \Phi \\ \nonumber && -\Big\lbrace \dfrac{1}{3} \theta ^{2} +\dfrac{1}{2(\phi+1)}\Big (\rho_{m}- (Rf'-f)+\theta \dot{\phi} +3\ddot{\phi}- \dfrac{3\dot{\phi}\phi^{\prime \cdot}}{\phi '} +\dfrac{3\phi'' \dot{\phi}^2}{\phi'^2}\Big)\Big\rbrace \dot{V}^{m}+\nabla^{2}\dot{V}^{m}=0\;, \\ && \label{dotd}
\dot{\Delta}^{m}+Z+\theta \mathcal{A} +\nabla^{2}V^{m}=0\;, \\&& \label{dotA}
\dot{\mathcal{A}}-\frac{1}{3}\theta \dot{V}^{m}+\frac{\rho_{m}}{2(1+\phi)}  V^{m} +\frac{1}{(1+\phi)} \dot{\Phi}+\frac{1}{2(1+\phi)}  \Big(\frac{1}{3}\theta - \frac{\dot{\phi}^{\prime}}{\phi^{\prime}}- \frac{2\dot{\phi}}{(1+\phi)}\Big)\Phi=0 \;, \\&& \label{dotphi}
\dot{\Phi}- \Psi-\dot{\phi}\mathcal{A}=0\;, \\&& \label{dotpsi}
\dot{\Psi}- \dfrac{\dddot{\phi}}{\dot{\phi}}\Phi- \ddot{\phi}\mathcal{A}=0\;,  \label{dotv}\\ && 
\dot{V}^{m}+\mathcal{A}=0\;,\\&&  \label{ddotdelta}
\ddot{\Delta}^{m}+ \Big( \dfrac{2}{3} \theta + \dfrac{\dot{\phi}}{2(\phi+1)}\Big)\dot{\Delta}^{m}- \dfrac{\rho_{m}}{2(\phi+1)}\Delta^{m} + \dfrac{\dot{\phi}}{2(\phi+1)}\tilde{ \nabla}^{2}V^{m} -\dfrac{\theta \rho_{m}}{2(\phi+1)}V^{m}  \nonumber\\ &&+
\Big \lbrace\dfrac{1}{3}\theta^{2} + \dfrac{1}{(\phi+1)}\Big(\rho_{m}- (Rf'-f)+3\ddot{\phi}- \dfrac{3\dot{\phi}\phi^{\prime \cdot}}{\phi '} +\dfrac{3\phi'' \dot{\phi}^2}{\phi'^2}\Big)\Big\rbrace\dot{V}^{m}\nonumber \\&& - \dfrac{3\theta}{2(\phi+1)} \dot{\Phi}+ \dfrac{1}{2(\phi+1)} \nabla^{2}\Phi+ \dfrac{1}{2(\phi+1)^{2}}\Big\lbrace 2\rho_{m} +(Rf'-f) \nonumber \\&&+3 (\phi+1)\Big(\frac{\ddot{\phi} \phi^{\prime \cdot}}{ \phi^{\prime} \dot{\phi}} - \frac{ \dddot{\phi}}{ \dot{\phi}} + \frac{ \phi^{\prime \cdot \cdot}}{ \phi^{\prime}} -\frac{ \phi^{\prime \cdot 2}}{\phi^{\prime 2}} - \frac{\phi ^{\prime \prime \cdot} \dot{\phi}}{\phi^{\prime2} }- \frac{2 \phi^{\prime \prime} \ddot{\phi}}{ \phi^{\prime 2}} + \frac{2 \phi^{\prime \prime} \phi^{\prime \cdot} \dot{\phi}}{ \phi^{\prime3}}+\dfrac{\theta\dot{\phi}^{\prime}}{3\phi'}- \dfrac{1}{9} \theta^{2}\Big) \Big\rbrace  \Phi=0\;,  \\&& \label{ddotphi}
\ddot{\Phi}+ \dfrac{\dot{\phi}}{(\phi+1)} \dot{\Phi}-\Big\lbrace \dfrac{\dddot{\phi}}{\dot{\phi}}-\dfrac{1}{2(\phi+1)}\Big( \dfrac{\theta \dot{\phi}}{3} - \dfrac{\dot{\phi} \dot{\phi}'}{\phi'} - \dfrac{2\dot{\phi}^{2}}{(\phi+1)}\Big) \Big\rbrace \Phi \\ \nonumber && + \Big\lbrace 2\ddot{\phi}- \dfrac{1}{3}  \theta \dot{\phi} \Big\rbrace \dot{V}^{m}+\dfrac{\rho_{m}\dot{\phi}}{2(\phi+1)}V^{m}=0\;, \label{ddotv}\\ &&
\ddot{V}^{m}+ \dfrac{1}{3}\theta  \dot{V}^{m} -\dfrac{\rho_{m}}{2(\phi+1)} V^{m} - \dfrac{1}{(1+\phi)} \dot{\Phi}-\dfrac{1}{2(\phi+1)}\Big( \dfrac{1}{3}\theta -\dfrac{\dot{\phi}'}{\phi'} -\dfrac{2 \dot{\phi}}{(\phi+1)}\Big)\Phi=0\;.
\end{eqnarray}
\subsection{Harmonic decomposition}
The above evolution Eqs. \eqref{dotz} -\eqref{ddotv}   can be thought of as a coupled system of harmonic
oscillator differential equations of the form \cite{gidelew2013beyond,carloni2006gauge}
\begin{eqnarray}
\ddot{X}+ A\dot{x}+ BX= C(Y,\dot{Y})\; , \label{Hdecomposition}
\end{eqnarray}
where $A$, $B$ and $C$ are independent of $X$ and they represent friction (damping), restoring and source forcing terms respectively.
To solve Eq. \eqref{Hdecomposition}, a separation of variables is applied such that
$$X(x,t)= X(\vec{x}) X(t), \hspace*{1cm} Y(x,t)= Y(\vec{x})Y(t)\; .$$
\newline
Since the evolution equations obtained so far are complicated to be solved, the harmonic decomposition approach is applied to these equations using the eigenfunctions and the corresponding wave number for these equations, therefore we write
$$X= \sum_{k}X^{k}Q_{k}(\vec{x})\; , \hspace*{1cm} Y=\sum_{k}Y^{k}(t)Q_{k}(\vec{x})\;,$$
where $Q_{k}(x)$ are the eigenfunctions of the covariantly defined spatial Laplace-Beltrami
operator \cite{gidelew2013beyond,carloni2006gauge}, such that
$$\tilde{\nabla}^{2}Q= -\dfrac{k^{2}}{a^{2}}Q\; .$$
The order of the harmonic (wave number) is given by $k=\dfrac{2\pi a}{\lambda}\; ,$ where $\lambda$ is the physical wavelength of the mode. The eigenfunctions $Q$ are covariantly constant, ie  $\dot{Q}_{k}(\vec{x})=0\; .$ Therefore, Eqs. \eqref{dotz}-\eqref{ddotv}  become 
\begin{eqnarray} \label{dotz1}
&&\dot{Z}_{k}+\Big(\dfrac{2}{3}\theta + \dfrac{\dot{\phi}}{2(\phi+1)}\Big)Z_{k}+ \dfrac{\rho_{m}}{2(\phi+1)}\Delta^{m}_{k}+\dfrac{\theta}{2(\phi+1)} \dot{\Phi}_{k}  \nonumber\\ && -\dfrac{1}{2(\phi+1)^{2}}\Big\lbrace2 \rho_{m} + (Rf'-f)-2\theta \dot{\phi} \nonumber \\&&+ 3(\phi+1)\Big(\frac{\ddot{\phi} \phi^{\prime \cdot}}{ \phi^{\prime} \dot{\phi}} - \frac{ \dddot{\phi}}{ \dot{\phi}} + \frac{ \phi^{\prime \cdot \cdot}}{ \phi^{\prime}} -\frac{ \phi^{\prime \cdot 2}}{\phi^{\prime 2}} - \frac{\phi ^{\prime \prime \cdot} \dot{\phi}}{\phi^{\prime2} }- \frac{2 \phi^{\prime \prime} \ddot{\phi}}{ \phi^{\prime 2}} + \frac{2 \phi^{\prime \prime} \phi^{\prime \cdot} \dot{\phi}}{ \phi^{\prime3}}
- \frac{k^2}{3a^2}\Big) \Big\rbrace \Phi_{k} \\ \nonumber && -\Big\lbrace \dfrac{1}{3} \theta ^{2} + \frac{k^2}{a^2}+\dfrac{1}{2(\phi+1)}\Big (\rho_{m}- (Rf'-f)+\theta \dot{\phi} +3\ddot{\phi}- \dfrac{3\dot{\phi}\phi^{\prime \cdot}}{\phi '} +\dfrac{3\phi'' \dot{\phi}^2}{\phi'^2}\Big)\Big\rbrace \dot{V}^{m}_{k}=0\;, \\ && \label{dotd}
\dot{\Delta}^{m}_k +Z_k - {\theta} \dot{V}^m_k - \dfrac{k^2}{a^2}V^{m}_k=0\;,  \\&& \label{dotA1}
\dot{\mathcal{A}}_{k}-\frac{1}{3}\theta \dot{V}^{m}_{k}+\frac{\rho_{m}}{2(1+\phi)}  V^{m}_{k} +\frac{1}{(1+\phi)} \dot{\Phi}_{k}+\frac{1}{2(1+\phi)}  \Big(\frac{1}{3}\theta - \frac{\dot{\phi}^{\prime}}{\phi^{\prime}}- \frac{2\dot{\phi}}{(1+\phi)}\Big)\Phi_{k}=0 \;, \\&& \label{dotphi}
\dot{\Phi}_{k}- \Psi-\dot{\phi}\mathcal{A}_{k}=0\;, \\&& \label{dotpsi1}
\dot{\Psi}_{k}- \dfrac{\dddot{\phi}}{\dot{\phi}}\Phi_{k}- \ddot{\phi}\mathcal{A}_{k}=0\;,  \label{dotv1}\\ && 
\dot{V}^{m}_{k}+\mathcal{A}_{k}=0\;,\\&&  \label{ddotdelta1}
\ddot{\Delta}^{m}_{k}+ \Big( \dfrac{2}{3} \theta + \dfrac{\dot{\phi}}{2(\phi+1)}\Big)\dot{\Delta}^{m}_{k}- \dfrac{\rho_{m}}{2(\phi+1)}\Delta^{m}_{k} - \dfrac{1}{2(\phi+1)}\Big(\theta \rho_{m}+ \frac{k^2 \dot{\phi}}{a^2}\Big)V^{m}_{k}  \nonumber\\ &&+
\Big \lbrace\dfrac{1}{3}\theta^{2} + \dfrac{1}{(\phi+1)}\Big(\rho_{m}- (Rf'-f)+3\ddot{\phi}- \dfrac{3\dot{\phi}\phi^{\prime \cdot}}{\phi '} +\dfrac{3\phi'' \dot{\phi}^2}{\phi'^2}\Big)\Big\rbrace\dot{V}^{m}_{k} - \dfrac{3\theta}{2(\phi+1)} \dot{\Phi}_{k} \nonumber\\ &&+ \dfrac{1}{2(\phi+1)^{2}}\Big\lbrace 2\rho_{m} +(Rf'-f) +3 (\phi+1)\Big(\frac{\ddot{\phi} \phi^{\prime \cdot}}{ \phi^{\prime} \dot{\phi}} - \frac{ \dddot{\phi}}{ \dot{\phi}} + \frac{ \phi^{\prime \cdot \cdot}}{ \phi^{\prime}} -\frac{ \phi^{\prime \cdot 2}}{\phi^{\prime 2}} \nonumber \\&&- \frac{\phi ^{\prime \prime \cdot} \dot{\phi}}{\phi^{\prime2} }- \frac{2 \phi^{\prime \prime} \ddot{\phi}}{ \phi^{\prime 2}} + \frac{2 \phi^{\prime \prime} \phi^{\prime \cdot} \dot{\phi}}{ \phi^{\prime3}}+\dfrac{\theta\dot{\phi}^{\prime}}{3\phi'}- \dfrac{1}{9} \theta^{2} -\frac{k^2}{3 a^2}\Big) \Big\rbrace  \Phi_{k}=0\;, 
\end{eqnarray}
\begin{eqnarray} \label{ddotphi1}
&&\ddot{\Phi}_{k}+ \dfrac{\dot{\phi}}{(\phi+1)} \dot{\Phi}_{k}-\Big\lbrace \dfrac{\dddot{\phi}}{\dot{\phi}}-\dfrac{1}{2(\phi+1)}\Big( \dfrac{\theta \dot{\phi}}{3} - \dfrac{\dot{\phi} \dot{\phi}'}{\phi'} - \dfrac{2\dot{\phi}^{2}}{(\phi+1)}\Big) \Big\rbrace \Phi _{k}\\ \nonumber && + \Big\lbrace 2\ddot{\phi}- \dfrac{1}{3}  \theta \dot{\phi} \Big\rbrace \dot{V}^{m}_{k}+\dfrac{\rho_{m}\dot{\phi}}{2(\phi+1)}V^{m}_{k}=0\;, \label{ddotv1}\\ &&
\ddot{V}^{m}_{k}+ \dfrac{1}{3}\theta  \dot{V}^{m}_{k} -\dfrac{\rho_{m}}{2(\phi+1)} V^{m}_{k} - \dfrac{1}{(1+\phi)} \dot{\Phi}_{k}-\dfrac{1}{2(\phi+1)}\Big( \dfrac{1}{3}\theta -\dfrac{\dot{\phi}'}{\phi'} -\dfrac{2 \dot{\phi}}{(\phi+1)}\Big)\Phi_{k}=0\;.
\end{eqnarray}
Then, we will study the growth of the matter density contrast with cosmological redshift. We apply a transformation to any time derivative function $f$ and H into a redshift derivative. Therefore, our evolution equations can be written as follow \footnote{ The harmonic decomposition is still applied but  from here onwards, $k$ will be removed from the following equations  to  avoid overcrowded notations.}:
\begin{eqnarray} \label{dotz11}
&&Z' - \frac{1}{(1+z)}\Big(\dfrac{2}{3}\theta + \dfrac{\dot{\phi}}{2(\phi+1)}\Big)Z- \dfrac{\rho_{m}}{2 H (1+z)(\phi+1)}\Delta_{m}+  \dfrac{\theta}{2(\phi+1)} \Phi'  \nonumber\\ && +\dfrac{1}{2H (1+z)(\phi+1)^{2}}\Big\lbrace2 \rho_{m} + (Rf'-f)-2\theta \dot{\phi}+ 3(\phi+1)\Big(\frac{\ddot{\phi} \phi^{\prime \cdot}}{ \phi^{\prime} \dot{\phi}} - \frac{ \dddot{\phi}}{ \dot{\phi}} \nonumber \\&&+ \frac{ \phi^{\prime \cdot \cdot}}{ \phi^{\prime}} -\frac{ \phi^{\prime \cdot 2}}{\phi^{\prime 2}} - \frac{\phi ^{\prime \prime \cdot} \dot{\phi}}{\phi^{\prime2} }- \frac{2 \phi^{\prime \prime} \ddot{\phi}}{ \phi^{\prime 2}} + \frac{2 \phi^{\prime \prime} \phi^{\prime \cdot} \dot{\phi}}{ \phi^{\prime3}}
- \frac{k^2}{3a^2}\Big) \Big\rbrace \Phi \\ \nonumber && - \Big\lbrace \dfrac{1}{3} \theta ^{2} + \frac{k^2}{a^2}+\dfrac{1}{2(\phi+1)}\Big (\rho_{m}- (Rf'-f)+\theta \dot{\phi} +3\ddot{\phi}- \dfrac{3\dot{\phi}\phi^{\prime \cdot}}{\phi '} +\dfrac{3\phi'' \dot{\phi}^2}{\phi'^2}\Big)\Big\rbrace V’_{m}=0\;, \\ && \label{dotd11}
 \Delta'_m -\frac{1}{H(1+z)}Z - {\theta} V'_m+ \dfrac{k^2}{a^2H (1+z)}V_{m}=0\;, \\&& \label{dotA1}
\mathcal{A}’-\frac{1}{3}\theta V’_{m}-\frac{\rho_{m}}{2H(1+\phi)(1+z)}  V_{m} +\frac{1}{(1+\phi)} \Phi’\nonumber\\&&-\frac{1}{2H(1+\phi)(1+z)}  \Big(\frac{1}{3}\theta - \frac{\dot{\phi}^{\prime}}{\phi^{\prime}}- \frac{2\dot{\phi}}{(1+\phi)}\Big)\Phi =0\;, \\&& \label{dotphi1}
\Phi’+ \frac{1}{H(1+z)}\Psi+\frac{\dot{\phi}}{H(1+z)}\mathcal{A}=0\;, \\&& \label{dotpsi1}
\Psi’+ \dfrac{\dddot{\phi}}{ H \dot{\phi}(1+z)}\Phi+ \frac{\ddot{\phi}}{H(1+z)}\mathcal{A}=0\;, \\ && \label{dotv11}
V’_{m}-\frac{1}{H(1+z)}\mathcal{A}=0 \;,
\end{eqnarray}
\begin{eqnarray}\label{ddotdelta11}
&&\Delta''_{m}- \frac{1}{2(1+z)}\Big( 1 + \dfrac{\dot{\phi}}{H(\phi+1)}\Big)\Delta'_{m}- \dfrac{\rho_{m}}{2H^{2} (1+z)^{2} (\phi+1)}\Delta_{m} \nonumber \\&&-
\frac{1}{H(1+z)}\Big \lbrace\dfrac{1}{3}\theta^{2} + \dfrac{1}{(\phi+1)}\Big(\rho_{m}- (Rf'-f)+3\ddot{\phi}- \dfrac{3\dot{\phi}\phi^{\prime \cdot}}{\phi '} +\dfrac{3\phi'' \dot{\phi}^2}{\phi'^2}\Big)\Big\rbrace V’_{m}\nonumber\\&&- \dfrac{1}{2H^{2} (1+z)^{2}(\phi+1)}\Big(\theta \rho_{m}+ \frac{k^2 \dot{\phi}}{a^2}\Big)V_{m} \nonumber\\&& +\dfrac{3\theta}{2H (1+z)(\phi+1)} \Phi' + \dfrac{1}{2H^{2}(1+z)^{2}(\phi+1)^{2}}\Big\lbrace 2\rho_{m} +(Rf'-f) \nonumber \\&&+3 (\phi+1)\Big(\frac{\ddot{\phi} \phi^{\prime \cdot}}{ \phi^{\prime} \dot{\phi}} - \frac{ \dddot{\phi}}{ \dot{\phi}} + \frac{ \phi^{\prime \cdot \cdot}}{ \phi^{\prime}} -\frac{ \phi^{\prime \cdot 2}}{\phi^{\prime 2}} - \frac{\phi ^{\prime \prime \cdot} \dot{\phi}}{\phi^{\prime2} }- \frac{2 \phi^{\prime \prime} \ddot{\phi}}{ \phi^{\prime 2}} + \frac{2 \phi^{\prime \prime} \phi^{\prime \cdot} \dot{\phi}}{ \phi^{\prime3}}\nonumber\\&&+\dfrac{\theta\dot{\phi}}{3\phi'}- \dfrac{1}{9} \theta^{2} -\frac{k^2}{3 a^2}\Big) \Big\rbrace  \Phi=0\;,\\
&&\Phi''+ \frac{1}{(1+z)} \Big(\frac{3}{2}-\dfrac{\dot{\phi}}{H(\phi+1)} \Big) \Phi’-\frac{1}{H^{2}(1+z)^{2}}\Big\lbrace \dfrac{\dddot{\phi}}{\dot{\phi}}-\dfrac{1}{2(\phi+1)}\Big( \dfrac{\theta \dot{\phi}}{3} - \dfrac{\dot{\phi} \dot{\phi}'}{\phi'}  \nonumber\\ &&- \dfrac{2\dot{\phi}^{2}}{(\phi+1)}\Big) \Big\rbrace \Phi  -\frac{1}{H(1+z)}\Big( 2\ddot{\phi}- \dfrac{1}{3}  \theta \dot{\phi} \Big) V’_{m}+ \dfrac{\rho_{m}\dot{\phi}}{2H^{2} (1+z)^{2}(\phi+1)}V_{m}=0\; ,  \label{ddotv11}\\ &&
V''_{m}+ \dfrac{1}{2(1+z)}V’_{m} -\dfrac{\rho_{m}}{2H^{2}(\phi+1)(1+z)^{2}} V_{m} + \dfrac{1}{H(1+\phi)(1+z)} \Phi’ \\ \nonumber &&-\dfrac{1}{2H^{2}(\phi+1)(1+z)^{2}}\Big( \dfrac{1}{3}\theta -\dfrac{\dot{\phi}'}{\phi'} -\dfrac{2 \dot{\phi}}{(\phi+1)}\Big)\Phi=0\;.
\end{eqnarray}
For more simplicity, we introduce here some quantities such as:
\begin{eqnarray}\label{q1}
&&\mathcal{B}_{1} =\dfrac{1}{(\phi+1)}\Big\lbrace2 \rho_{m} + (Rf'-f)-2\theta \dot{\phi}+ 3(\phi+1)\Big(\frac{\ddot{\phi} \phi^{\prime \cdot}}{ \phi^{\prime} \dot{\phi}} - \frac{ \dddot{\phi}}{ \dot{\phi}} \nonumber \\&&+ \frac{ \phi^{\prime \cdot \cdot}}{ \phi^{\prime}} -\frac{ \phi^{\prime \cdot 2}}{\phi^{\prime 2}} - \frac{\phi ^{\prime \prime \cdot} \dot{\phi}}{\phi^{\prime2} }- \frac{2 \phi^{\prime \prime} \ddot{\phi}}{ \phi^{\prime 2}} + \frac{2 \phi^{\prime \prime} \phi^{\prime \cdot} \dot{\phi}}{ \phi^{\prime3}}
- \frac{k^2}{3a^2}\Big) \Big\rbrace\;, \\&&
\mathcal{B}_{2}= \Big\lbrace \dfrac{1}{3} \theta ^{2} + \frac{k^2}{a^2}+\dfrac{1}{2(\phi+1)}\Big (\rho_{m}- (Rf'-f)+\theta \dot{\phi} +3\ddot{\phi}- \dfrac{3\dot{\phi}\phi^{\prime \cdot}}{\phi '} +\dfrac{3\phi'' \dot{\phi}^2}{\phi'^2}\Big)\Big\rbrace\;, \\&&
\mathcal{B}_{3}= \Big \lbrace\dfrac{1}{3}\theta^{2} + \dfrac{1}{(\phi+1)}\Big(\rho_{m}- (Rf'-f)+3\ddot{\phi}- \dfrac{3\dot{\phi}\phi^{\prime \cdot}}{\phi '} +\dfrac{3\phi'' \dot{\phi}^2}{\phi'^2}\Big)\Big\rbrace\;, \\&&
\mathcal{B}_{4}= \dfrac{1}{(\phi+1)}\Big\lbrace 2\rho_{m} +(Rf'-f) + 3(\phi+1)\Big(\frac{\ddot{\phi} \phi^{\prime \cdot}}{ \phi^{\prime} \dot{\phi}} - \frac{ \dddot{\phi}}{ \dot{\phi}} \nonumber \\&&+ \frac{ \phi^{\prime \cdot \cdot}}{ \phi^{\prime}} -\frac{ \phi^{\prime \cdot 2}}{\phi^{\prime 2}} - \frac{\phi ^{\prime \prime \cdot} \dot{\phi}}{\phi^{\prime2} }- \frac{2 \phi^{\prime \prime} \ddot{\phi}}{ \phi^{\prime 2}} + \frac{2 \phi^{\prime \prime} \phi^{\prime \cdot} \dot{\phi}}{ \phi^{\prime3}}+\dfrac{\theta\dot{\phi}^{\prime}}{3\phi'}- \dfrac{1}{9} \theta^{2} -\frac{k^2}{3 a^2}\Big) \Big\rbrace \;, \\ &&
\mathcal{B}_{5}= \Big\lbrace \dfrac{\dddot{\phi}}{\dot{\phi}}-\dfrac{1}{2(\phi+1)}\Big( \dfrac{\theta \dot{\phi}}{3} - \dfrac{\dot{\phi} \dot{\phi}'}{\phi'} - \dfrac{2\dot{\phi}^{2}}{(\phi+1)}\Big) \Big\rbrace \;, \\&& \label{q6}
\mathcal{B}_{6}= \Big( \dfrac{1}{3}\theta -\dfrac{\dot{\phi}'}{\phi'} -\dfrac{2 \dot{\phi}}{(\phi+1)}\Big)\;.
\end{eqnarray}
By using the introduced dimensionless variables from Eq. \eqref{e4}, then we rewrite the more generalised form of the evolution equations as follows:
\begin{eqnarray} \label{dotz111}
&&Z' - \frac{1}{(1+z)}\Big(2+ \dfrac{\mathcal{Y}}{2}\Big)Z- \dfrac{3H \tilde{\Omega}_{m}}{2 (1+z)}\Delta_{m}+  \dfrac{3H}{2(\phi+1)} \Phi' \nonumber \\&&+\dfrac{\mathcal{B}_{1}}{2H (1+z)(\phi+1)} \Phi - \mathcal{B}_{2} V’_{m}=0\;, \\ && \label{dotd111}
 \Delta’_m-\frac{1}{H(1+z)}Z - 3HV’_m+ \dfrac{k^2}{a^2 	H (1+z)}V_{m}=0\;, \\&& \label{dotA111}
\mathcal{A}'-HV'_{m}-\frac{3H \tilde{\Omega}_{m}}{2(1+z)}  V_{m} +\frac{1}{(1+\phi)} \Phi'-\frac{1}{2(1+\phi)(1+z)}  \Big(1 - \frac{\dot{\phi}^{\prime}}{H\phi^{\prime}}- 2\mathcal{Y}\Big)\Phi =0\;, \\&& \label{dotphi111}
\Phi’+ \frac{1}{H(1+z)}\Psi+\frac{\dot{\phi}}{H(1+z)}\mathcal{A}=0\;, \\&& \label{dotpsi111}
\Psi’+ \dfrac{\dddot{\phi}}{ H \dot{\phi}(1+z)}\Phi+ \frac{\ddot{\phi}}{H(1+z)}\mathcal{A}=0\;,  \label{dotv111}\\ && 
V’_{m}-\frac{1}{H(1+z)}\mathcal{A}=0 \;, \\&& \label{ddotdelta111}
\Delta''_{m}- \frac{1}{2(1+z)}\Big( 1 + \mathcal{Y}\Big)\Delta'_{m}- \dfrac{3\tilde{\Omega}_{m}}{2 (1+z)^{2}}\Delta_{m} - \dfrac{H}{2(1+z)^{2}}\Big(9\tilde{\Omega}_{m}+ \frac{k^2 \mathcal{Y}}{a^{2} H^{2}}\Big)V_{m} \nonumber \\&&-
\frac{\mathcal{B}_{3}}{H(1+z)}  V’_{m}   +\dfrac{9}{2 (1+z)(\phi+1)} \Phi' + \dfrac{\mathcal{B}_{4}}{2H^{2}(1+z)^{2}(\phi+1)}  \Phi=0\;,
\\&& \label{ddotphi111}
\Phi''+ \frac{1}{(1+z)} \Big(\frac{3}{2}-\mathcal{Y} \Big) \Phi’-\frac{\mathcal{B}_{5}}{H^{2}(1+z)^{2}} \Phi  -\frac{\dot{\phi}}{(1+z)}\Big( \frac{2\ddot{\phi}}{H\dot{\phi}}- 1\Big) V’_{m}\nonumber \\&&+ \dfrac{3\tilde{\Omega}_{m}\dot{\phi}}{2 (1+z)^{2}}V_{m}=0\; , \\ && \label{ddotv11}
V''_{m}+ \dfrac{1}{2(1+z)}V’_{m} -\dfrac{3\tilde{\Omega}_{m}}{2(1+z)^{2}} V_{m} + \dfrac{1}{H(1+\phi)(1+z)} \Phi’\nonumber \\&&-\dfrac{\mathcal{B}_{6}}{2H^{2}(\phi+1)(1+z)^{2}}\Phi=0\;.
\end{eqnarray}
For further analysis, in this part we are going to apply the quasi-static approximation to our evolution equations  Eqs. \eqref{ddotdelta111} - \eqref{ddotv11}.  In this approximation, terms involving time derivatives for gravitational potential are neglected and only those terms involving density perturbation are kept. Therefore, $ \Phi''= \Phi'=0$. Eqs. \eqref{ddotdelta111} - \eqref{ddotv11} become
\begin{eqnarray}\label{phi12}
&& \Phi=- \dfrac{ H^2 \dot{\phi}(1+z)  \Big(\frac{2\ddot{\phi}}{H\dot{\phi}}- 1\Big) }{\mathcal{B}_{5}} V’_{m} + \dfrac{3 	H^2 \dot{\phi} \tilde{\Omega}_{m}}{2\mathcal{B}_{5}} V_{m}\;, \\ \label{v12}
&& V''_{m}+\dfrac{1}{2(1+z)}\Big\lbrace 1+ \dfrac{H\mathcal{Y} \mathcal{B}_{6}  \Big(\frac{2\ddot{\phi}}{H\dot{\phi}}- 1\Big)}{\mathcal{B}_{5}}\Big \rbrace V’_{m} - \dfrac{3\tilde{\Omega}_{m}}{2(1+z)^2}\Big\lbrace 1+ \dfrac{H\mathcal{Y} \mathcal{B}_{6}}{2 \mathcal{B}_{5}}\Big \rbrace V_{m}=0\;, \\ \label{D12}
&&\Delta''_{m}- \frac{1}{2(1+z)}\Big( 1 + \mathcal{Y}\Big)\Delta’_{m}- \dfrac{3\tilde{\Omega}_{m}}{2 (1+z)^{2}}\Delta_{m} - \frac{1}{H(1+z)} \Big\lbrace \mathcal{B}_{3} + \frac{H^{2} \mathcal{Y} \mathcal{B}_{4} \Big(\frac{2\ddot{\phi}}{H\dot{\phi}}- 1\Big) }{2 \mathcal{B}_{5}}\Big\rbrace V’_{m}\nonumber \\&&- \frac{H}{2(1+z)} \Big( 9\tilde{\Omega}_{m} + \frac{k^{2} \mathcal{Y}}{a^{2} H^{2}} -\frac{3 \tilde{\Omega}_{m} \mathcal{Y} \mathcal{B}_{4}}{2\mathcal{B}_{5}} \Big) V_{m}=0\;.
\end{eqnarray}
The  above evolution Eqs.  \eqref{dotz111} - \eqref{ddotv11}  are exactly the same  as the work presented in \cite{maartens98} for GR. The GR can be recovered for the case of $f(R) = R$, therefore $(\phi +1)=1 $ and we have 
\begin{eqnarray}
&& Z' - \dfrac{2}{(1+z)}Z -\dfrac{3 H\tilde{\Omega}_{m}}{2(1+z)} \Delta_m  +3H^2 \bigg(-1 -\frac{\tilde{\Omega}_{m}}{2}- \dfrac{k^2}{3 H^2 a^2} \bigg) V’_m=0\;, 
 \\&&   \Delta’_m -\frac{1}{H(1+z)}Z- {\theta} V’_m+ \dfrac{k^2}{a^2 	H (1+z)}V_{m}=0\label{matterdenityfluid0GR}\;, \\
 &&\mathcal{A}’-H V’_{m} - \dfrac{3 H\tilde{\Omega}_{m}}{2 (1+z)}V_m=0  \label{accelaration0GR}\;, \\&&
V’_{m}- \frac{1}{H(1+z)}\mathcal{A}=0 \label{ddot1010000GR}\;, \\&& 
  \Delta''_m  - \frac{1}{2(1+z)}{\Delta}’_m - \frac{3\bar{\Omega}_m }{2(1+z)^2}\Delta_m-\frac{3H}{(1+z)}\Big(1+\bar{\Omega}_m \Big)V’_m- \frac{9\bar{\Omega}_mH}{2(1+z)^2}V_m= 0 \label{GRdensity1}\;,\\&&
 V''_{m}   + \frac{1}{2(1+z)}{V}’_m  -  \dfrac{3{\Omega}_m }{2(1+z)^2} V_m = 0\;.\label{GRV}
\end{eqnarray}
In the following section, we explore  the solutions of the density and velocity contrast in GR and  one of the $f(R)$ toy models.
\section{Solutions}\label{sec7}
In this section we will solve the whole system of  perturbations equations we obtained so far  Eqs. \eqref{dotz111} - \eqref{ddotv11}  to explore the growth of the matter density contrast in the GR context and for $f(R)$ as a scalar-tensor theory of gravity for non-quasi-static approximations and quasi-static approximation from Eqs. \eqref{phi12} - \eqref{D12}. The exact solutions of the matter density contrast can be found in the quasi static approximations and the numerical solution will be presented in the non-quasi-static approximations as well.  We defined the normalized energy density for matter fluid as
 \begin{equation}
\delta(z)=\frac{\Delta _m(z)}{\Delta (z_{in})} \;,                                                                                                                     
\end{equation}
where $\Delta (z_{in})$ is the initial value of $ \Delta_{m}(z)$ at $z_{in}$. In the same manner, we defind normalized velocity contrast as
\begin{equation}
\nu(z)=\frac{V_m (z)}{V(z_{in})} \;,                                                                                                                     
\end{equation}
\subsection{The growth of the velocity and the matter -density fluctuations in GR limits}
Here, we analyze the growth of matter energy density contrasts $\delta(z)$  Eq.  \eqref{GRdensity1} and the velocity contrast $\nu(z)$ Eq.  \eqref{GRV}  with cosmic-time.  The exact solutions of the velocity contrast  yields  as
\begin{equation}
 V_m (z) = c_1\left(1+z\right)^{-1}+ c_2\left(1+z\right)^{\frac{3}{2}}\;.\label{solutionVGR}
\end{equation}
The integration constant $c_1$ and $c_2$  can be determined by the imposing initial conditions for plotting.
Those constants are worthy to find the exact solution for the density contrasts, and  we present  in the following as
\begin{eqnarray}
&& {\it c_1}=\frac{-2}{5} (1+z_{in}) \Big((1+z_{in}) \dot{V}_m(z_{in}) -\dfrac{3}{2} V_m(z_{in})\Big)\nonumber\\&&{\it c_2}= \dfrac{2}{5 \sqrt{(1+z_{in}) }} \Big(\dot{V}_m(z_{in}) + \dfrac{V_m(z_{in})}{(1+z_{in})}\Big) \;.
\end{eqnarray}
Consequently, the second-order evolution equation of the matter density equation \eqref{GRdensity1} becomes a closed system by substituting the solutions of the velocity contrast Eq. \eqref{solutionVGR} and it's first order derivative.  Then, the exact solution is given as
\begin{eqnarray}
 && \Delta_m = \dfrac{1}{2} {\it c_{1}} (1+z)^{1/2}+3 {\it c_{2}} (1+z)^{3}+{\it c_{3}} (1+z)^{(3/4- \sqrt{33}/4)} + {\it c_{4}} (1+z)^{(3/4+ \sqrt{33}/4)}
  \;,\label{solutionGRd}
\end{eqnarray}
where $c_3$ and $c_4$ are the integration constants and they are given as
\begin{eqnarray}
&&{ \it c_{3, 4}}=\frac{\mp \frac{2}{\sqrt{33}}}{ (1+z_{in})^{\big(\frac{-1}{4} -\frac{\sqrt{33}}{4}\big)} } \Big\lbrace \dot{\Delta}_m(z_{in}) - \dfrac{\big(\frac{3}{4} \pm \frac{\sqrt{33}}{4}\big)}{(1+z_{in})} \Delta_m(z_in) \nonumber\\&&- \dfrac{ {\it c_{2}}\big(\frac{5}{4} \mp\frac{\sqrt{33}}{4}\big)}{2\sqrt{(1+z_{in})}} - 3{\it c_{1}} \Big(\frac{9}{4} \mp \frac{\sqrt{33}}{4}\Big) (1+z_{in})^{2}\Big\rbrace \;.
\end{eqnarray}
In the following figures we present  growth the matter density and velocity fluctuations with cosmological redshift in the GR approach. We set the initial conditions at $ V _{in}= V(z_{in}\simeq 1100)= 10^{-5}$ and $\dot{V}_{in}=\dot{V} (z_{in}=1100)= 0$ and $ \Delta _{in}= \Delta _{m}(z_{in}\simeq 1100)= 10^{-5}$ and $\dot{\Delta}_{in}=\dot{\Delta}_{m} (z_{in}=1100)= 0$. 
 \begin{figure}[H]
 \begin{minipage}{0.5\textwidth}
\includegraphics[width=0.9\textwidth]{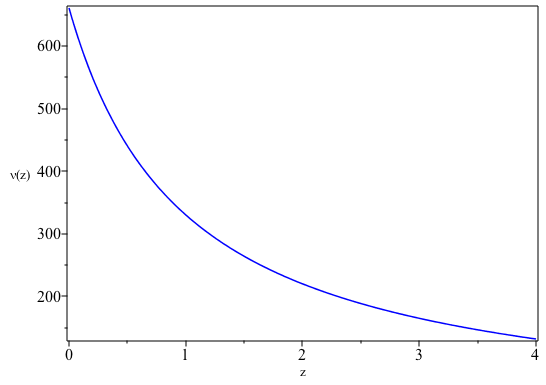}
    \caption{The growth of the velocity contrast  for Eq. \eqref{solutionVGR} (GR limits).}
    \label{fig:GRV}
\end{minipage} 
\qquad
 \begin{minipage}{0.5\textwidth}
\includegraphics[width=0.9\textwidth]{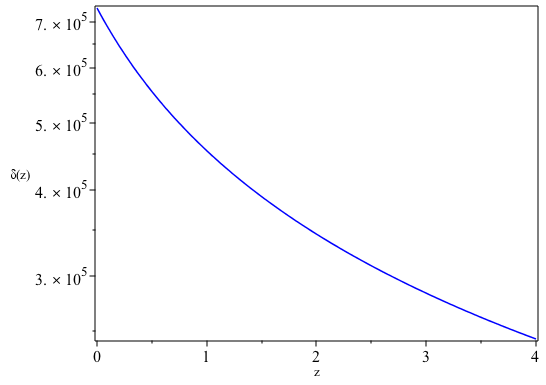}
   \caption{The growth of the density contrast for Eq. \eqref{solutionGRd} (GR limits).}
    \label{fig:GRd}
\end{minipage} 
 \end{figure}
From this plots, we depict clearly the contribution of dust component of the universe for the fluctuations of matter density and velocity  are growing  with decreasing red shift.
\subsection{The growth of the velocity and the matter -density fluctuations in $f(R)$ toy models}
In this sub-section, we consider $R^{n}$ model, one of the $f(R)$ toy models that are considered to be the simplest and widely studied form of higher order $f(R)$ gravitational theories.
The Lagrangian density of such models is given as
\begin{equation}
f(R)= \beta R^{n}\;,
\end{equation} 
where $\beta$ represents the coupling parameter and an arbitrary constant $n\neq1$ is considered for exploring cosmological models. In \cite{carloni2005cosmological}, it has been shown, using the cosmological dynamical systems approach, that the scale factor $a(t)$ admits an exact solution of the form
\begin{equation}
a= a_{0}(t/t_0)^{\frac{2n}{3(1+w)}}\;,
\end{equation}
with $w=0$ and normalized coefficients $\beta$ and $a_{0}$. One can obtain the following expressions for the expansion, the Ricci scalar and the effective matter energy density respectively:
\begin{eqnarray}
&&\theta= \frac{2n}{t}\; , \hspace*{.3cm} R=\frac{4n(4n-3)}{3t^{2}}\; ,\\
&&
\rho_{m}= n\Big(\frac{3}{4}\Big)^{1-n} \Big(\frac{4n^{2}-3n}{t^{2}}\Big)^{n-1}\Big(\frac{-16n^{2}+26n-6}{3t^{2}}\Big)\;.
\end{eqnarray}
Where  $t= t_{0} (1+z)^{-\frac{3}{2n}}$ and the Hubble parameter is $H(z)= \dfrac{1}{H_{0}} h(z)$, where  $H_{0} = t_{0}$ and $ h(z)= \dfrac{2n}{3} (1+z)^{3/2n}$.
 The definitions in Eqs. \eqref{q1}-\eqref{q6}, become
 \begin{eqnarray}\label{q11}
&&\mathcal{B}_{1} = H^{2} \Big( 6\tilde{\Omega}_{m} +6\mathcal{X} -6\mathcal{Y} + \frac{27(n-2) (1-2n)}{2n^{2}} - \frac{k^{2}}{ a^{2} H^{2}}\Big), \\&&
\mathcal{B}_{2}= H^{2} \Big( 3+ \frac{k^{2}}{ a^{2} H^{2}} + \frac{3\tilde{\Omega}_{m}}{2} -3\mathcal{X} + \frac{3}{2} \mathcal{Y} +\frac{27(n-1) (2n-1)}{4n^{2}}\Big)\;, \\&&
\mathcal{B}_{3}= H^{2} \Big( 3+ 3\tilde{\Omega}_{m} -6\mathcal{X} +\frac{27(n-1) (2n-1)}{2n^{2}}\Big)\;, \\&&
\mathcal{B}_{4}= H^{2} \Big( 6\tilde{\Omega}_{m} + 6\mathcal{X}  +\frac{3(1-2n)(5n-9)}{n^{2}}-1 - \frac{k^{2}}{a^{2} H^{2}}\Big) \;, \\ &&
\mathcal{B}_{5}=  \frac{3H^{2}}{4n^{2}} \Big( 32n^{2} -41 n+15\Big) \;, \\&& \label{q66}
\mathcal{B}_{6}=  \frac{H}{2n}\Big(20n -15\Big)\;.
\end{eqnarray}
 Where
 \begin{eqnarray}\label{0}
 && \mathcal{Y}= \dfrac{-3(n-1)}{n}\;, \hspace{.3cm} \label{1}
 \mathcal{X}= \dfrac{(n-1)(4n-3)}{2n^2}\;,\\&& \label{2} 
  (\phi+1)= n\Big(\dfrac{4n(4n-3)}{3}\Big)^{(n-1)} (1+z)^{(\frac{3(n-1)}{n})}\;, \\&&\label{6}
 \dot{\phi}= \dfrac{-2n(n-1)}{t_{0}}\Big(\dfrac{4n(42n-3)}{3}\Big)^{(n-1)} (1+z)^{(3-3/2n)}\;.
 \end{eqnarray}
 We redefine the following normalized quantity as:
\begin{eqnarray}
&&V_{m}= H_{0} v_{m}\;, \hspace{.3cm} k^{2}= \frac{1}{H^{2}_{0}} {K}^{2}\;,\hspace{.3cm}
Z= H_{0} \mathcal{Z}\;, \hspace{.3cm} \Psi= H_{0} \xi\;,
\end{eqnarray}
where $H(z)= \dfrac{1}{H_{0}} h(z)$, $H_{0}= t_{0}$, therefore Eqs.\eqref{dotz111} -  \eqref{ddotv11} become
\begin{eqnarray}\label{dotz12}
&&\mathcal{Z}’ - \frac{1}{(1+z)}\Big(2+ \dfrac{\mathcal{Y}}{2}\Big)\mathcal{Z}- \dfrac{3h(z) \tilde{\Omega}_{m}}{2 (1+z)}\Delta^{m}+  \dfrac{3h(z)}{2(\phi+1)} \Phi’\nonumber \\&&+\dfrac{h(z)}{2(1+z)(\phi+1)} \Big( 6\tilde{\Omega}_{m} +6\mathcal{X} -6\mathcal{Y} + \frac{27(n-2) (1-2n)}{2n^{2}} - \frac{k^{2}}{ a^{2} h(z)^{2}}\Big)\Phi\nonumber \\&& -h(z)^{2} \Big( 3+ \frac{k^{2}}{ a^{2} h(z)^{2}} + \frac{3\tilde{\Omega}_{m}}{2} -3\mathcal{X} + \frac{3}{2} \mathcal{Y} +\frac{27(n-1) (2n-1)}{4n^{2}}\Big) v’_{m}=0\;, \\ && \label{dotd111}
 \Delta’_m-\frac{1}{h(z)(1+z)}\mathcal{Z}- 3Hv’_m+ \dfrac{k^2}{a^2 	h(z) (1+z)}v_{m}=0\;, \\&& \label{dotA1112}
\mathcal{A}’-h(z)v’_{m}-\frac{3h(z) \tilde{\Omega}_{m}}{2(1+z)}  v_{m} +\frac{1}{(1+\phi)} \Phi’\nonumber \\&&-\frac{1}{2(1+\phi)(1+z)}  \Big(1 - \frac{3(1-2n)}{2n}- 2\mathcal{Y}\Big)\Phi =0\;, \\&& \label{dotphi1112}
\Phi’+ \frac{1}{h(z)(1+z)}\xi+\frac{\mathcal{Y} (\phi+1)}{(1+z)}\mathcal{A}=0\;, \\&& \label{dotpsi1112}
\xi'+ \dfrac{9(2n-1) h(z)}{ 2n(1+z)}\Phi+ n(-2n+2) (1-2n)\Big(\dfrac{4n(42n-3)}{3}\Big)^{(n-1)}(1+z)^{2} \mathcal{A}=0\;,  \label{dotv1112}\\ && 
v’_{m}-\frac{1}{h(z)(1+z)}\mathcal{A}=0 \;, \\&& \label{ddotdelta44}
 \Delta''_{m}-\dfrac{1}{2(1+z)} \Big(1+\mathcal{Y}\Big)\Delta’_{m}\nonumber \\&&-\dfrac{3\tilde{\Omega}_{m}}{2(1+z)^2} \Delta_{m} -\dfrac{h(z)}{(1+z)} \Big( 3 +3 \tilde{\Omega}_{m} -6 \mathcal{X} + \dfrac{27(n-1)(2n-1)}{2n^2}\Big) v’_{m}  \nonumber\\ && - \dfrac{h(z)}{2 (1+z)^2}\Big( 9 \tilde{\Omega}_{m} +\dfrac{\mathcal{Y} K^2}{a^2 h(z)^2}\Big) v_{m}+\dfrac{9}{2(\phi+1) (1+z)} \Phi’\nonumber \\&& + \dfrac{1}{2 (\phi+1) (1+z)^2}\Big( 6 \tilde{\Omega}_{m} +6 \mathcal{X} +\frac{3(1-2n)(5n-9)}{n^{2}}-1 -\frac{K^{2}}{a^{2} h(z)^2}\Big) \Phi=0\;, \\ &&\label{ddot phi44}
\Phi'' +\dfrac{1}{(1+z)} \Big( \dfrac{3}{2}-\mathcal{Y}\Big)\Phi' -\dfrac{3( 32n^{2}-41n+15)}{4n^{2}(1+z)^2}\Phi -\dfrac{( 3-7n)\dot{\phi}}{n(1+z)} v’ _{m}\nonumber\\&&+\dfrac{3\tilde{\Omega}_{m}\dot{\phi}}{2(1+z)^2} v_{m}=0\;, \\&&\label{ddotv44}
v''_{m} +\dfrac{1}{2(1+z)} v’_{m} -\dfrac{3\tilde{\Omega}_{m}}{2(1+z)^2}v_{m} + \dfrac{1}{h(z)(\phi+1) (1+z)}\Phi' \nonumber\\&&- \dfrac{(20n-15)}{4nh(z)(\phi+1) (1+z)^2} \Phi =0\;. 
\end{eqnarray}
In the following, we will solve the whole system of  perturbation equations, we start by solving the whole system of the first-order evolution equations \eqref{dotz12}- \eqref{dotv1112}. We have evaluated the numerical solutions simultaneously to analyze the density fluctuations for the short-wavelength modes, i.e. large values of the wave number $K$ and for the long-wavelength modes, i.e. small values of the wave number $K$.  We set the initial conditions at $ v _{in}= v(z_{in}\simeq 1100)= 10^{-5}$,  $ \Phi_{in}= \Phi(z_{in}\simeq 1100)= 10^-5$ , $\xi _{in}= \xi(z_{in}\simeq 1100)=10^{-5}$, $\mathcal{A} _{in}= \mathcal{A}(z_{in}\simeq 1100)= 10^{-5}$,  $\mathcal{Z} _{in}=\mathcal{ Z}(z_{in}\simeq 1100)= 10^{-5}$  and $\Delta _{in}= \Delta _{m}(z_{in}\simeq 1100)= 10^{-5}$. Then,  we  evaluated the numerical solutions from the second-order evolution  Eqs.\eqref{ddotdelta44} - \eqref{ddotv44}  simultaneously to analyze the density and velocity fluctuations with redshift for the short-wavelength modes and for the long-wavelength modes. For further analysis, we  analyse  the growth of the velocity and the matter density fluctuations with redshift and present the numerical results for different initial conditions as presented in \cite{abebe2013large}, to see how sensitive the results are to change the initial conditions.
\begin{itemize}
\item[I] $\Delta_{in}= v_{in}= \Phi_{in}= 10^{-5} $ and $\Delta’ = v’= \Phi’= 10^{-5}\;.$
\item[II] $\Delta_{in}= v_{in}= \Phi_{in}= 10^{-5} $ and $\Delta’ = v’= \Phi’= 0\;.$
\end{itemize}
\subsubsection{Case I: Solving the whole system }
\begin{figure}[H]
 \begin{minipage}{0.5\textwidth}
\includegraphics[width=0.9\textwidth]{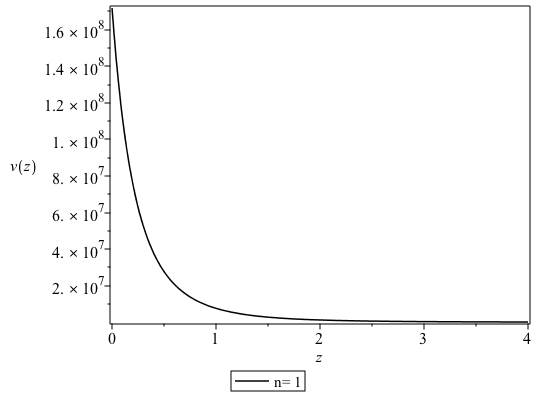}
    \caption{The growth of the velocity contrast versus cosmological redshift  for the system Eqs.  \eqref{dotz12}- \eqref{dotv1112}
    for $n = 1$ (GR).}
    \label{fig1f}
\end{minipage} 
\qquad
 \begin{minipage}{0.5\textwidth}
\includegraphics[width=0.9\textwidth]{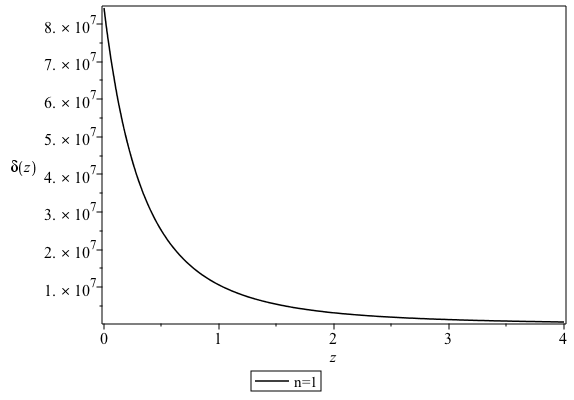}
   \caption{The growth of the density contrast versus cosmological redshift for  the system Eqs.  \eqref{dotz12}- \eqref{dotv1112} for $n = 1$ (GR).}
    \label{fig2f}
\end{minipage} 
\end{figure}

\begin{figure}[H]
\begin{minipage}{0.5\textwidth}
\includegraphics[width=0.9\textwidth]{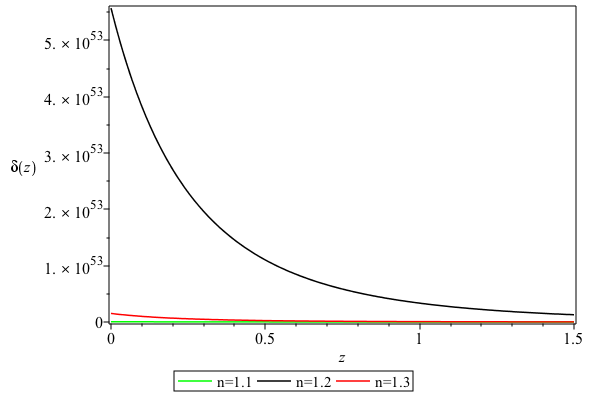}
    \caption{The growth of the density contrast versus cosmological redshift for the system of Eqs. \eqref{dotz12}- \eqref{dotv1112} for $n >1$ for long-wavelength ($K= 0.001$).}
    \label{Figure1f}
\end{minipage} 
\qquad
 \begin{minipage}{0.5\textwidth}
\includegraphics[width=0.9\textwidth]{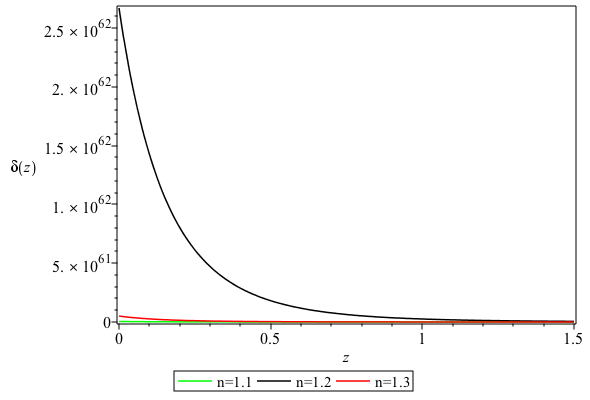}
   \caption{The growth of  the density contrast versus cosmological redshift for the system of Eqs.  \eqref{dotz12}- \eqref{dotv1112} for $n >1$  for short-wavelength ($K= 10^5$).}
    \label{Figure2f}
\end{minipage} 
\qquad
\begin{minipage}{0.5\textwidth}
\includegraphics[width=0.9\textwidth]{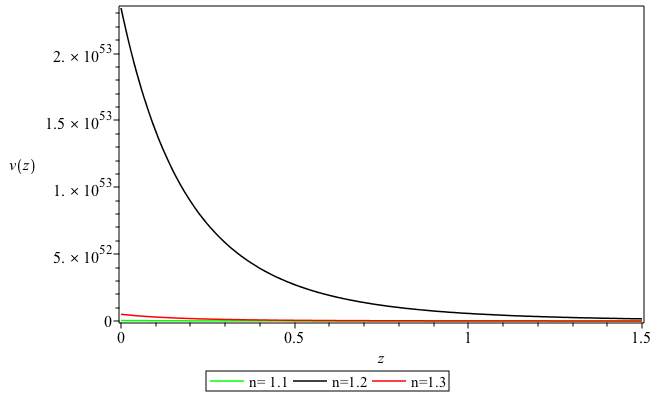}
   \caption{The growth of the velocity contrast versus cosmological redshift for the system of Eqs.  \eqref{dotz12}- \eqref{dotv1112} for $n >1$.}
    \label{Figure3f}
\end{minipage} 
\end{figure}

\begin{figure}[H]
 \begin{minipage}{0.5\textwidth}
\includegraphics[width=0.9\textwidth]{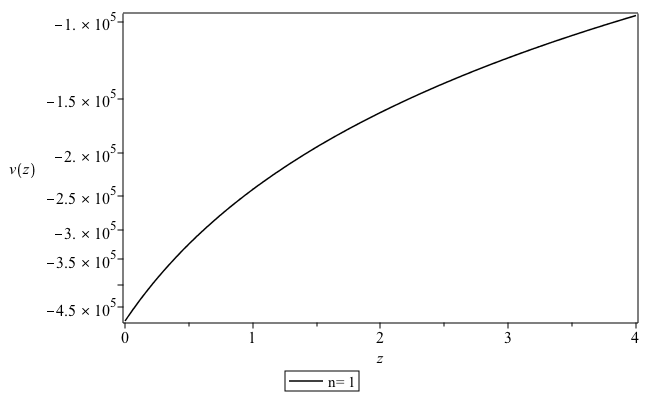}
    \caption{The growth of the velocity contrast versus cosmological redshift  for the system Eqs. \eqref{ddotdelta44}- \eqref{ddotv44}
    for $n = 1$ (GR) for set I.}
    \label{fig1}
\end{minipage} 
\qquad
 \begin{minipage}{0.5\textwidth}
\includegraphics[width=0.9\textwidth]{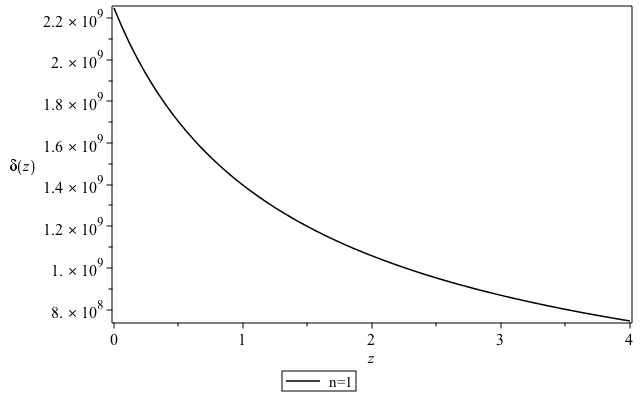}
   \caption{The growth of the density contrast versus cosmological redshift for  the system Eqs. \eqref{ddotdelta44}- \eqref{ddotv44} for $n = 1$ (GR) for set I.}
    \label{fig2}
\end{minipage} 
 \begin{minipage}{0.5\textwidth}
\includegraphics[width=0.9\textwidth]{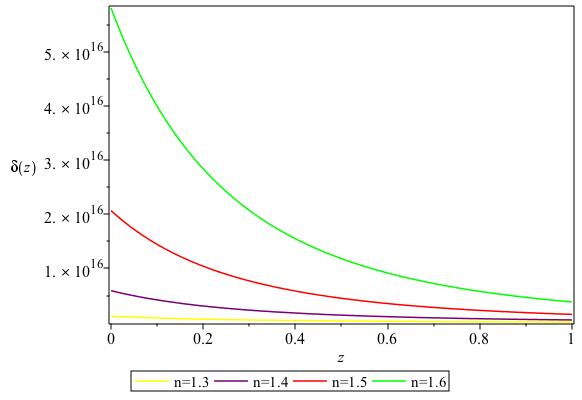}
    \caption{The growth of the density contrast versus cosmological redshift for the system of Eqs. \eqref{ddotdelta44}- \eqref{ddotv44} for $n >1$ for long-wavelength ($K= 0.001$) for set I.}
    \label{Figure1}
\end{minipage} 
\qquad
 \begin{minipage}{0.5\textwidth}
\includegraphics[width=0.9\textwidth]{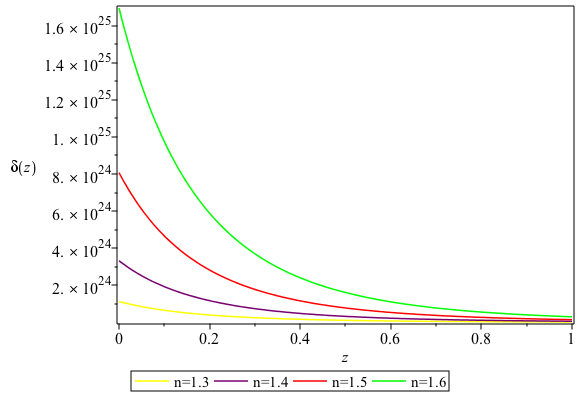}
   \caption{The growth of  the density contrast versus cosmological redshift for the system of Eqs. \eqref{ddotdelta44}- \eqref{ddotv44} for $n >1$  for short-wavelength ($K= 10^5$) for set I.}
    \label{Figure2}
\end{minipage} 
\qquad
\begin{minipage}{0.5\textwidth}
\includegraphics[width=0.9\textwidth]{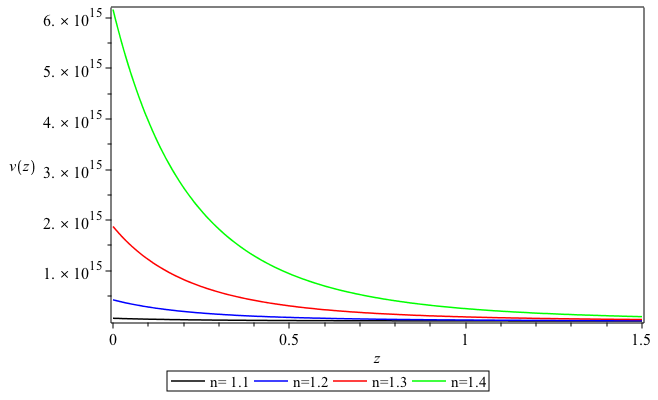}
    \caption{The growth of  the velocity contrast versus cosmological redshift for the system of Eqs. \eqref{ddotdelta44}- \eqref{ddotv44} for $n >1$  for set I.}
    \label{Figure3}
\end{minipage} 
\qquad
\begin{minipage}{0.5\textwidth}
\includegraphics[width=0.9\textwidth]{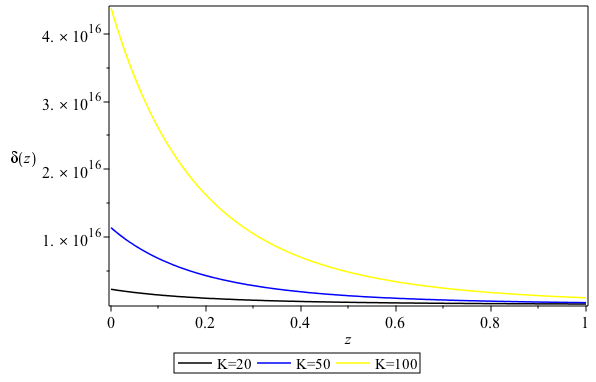}
    \caption{The growth of the density contrast versus cosmological redshift for the system of Eqs. \eqref{ddotdelta44}- \eqref{ddotv44} for $n=1.1$  for set I and different values of $K$.}
    \label{Figure4}
\end{minipage} 
\end{figure}

\begin{figure}[H]
  \begin{minipage}{0.5\textwidth}
\includegraphics[width=0.9\textwidth]{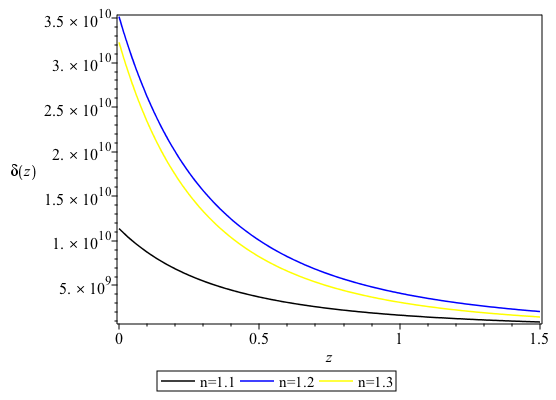}
    \caption{The growth of the  density contrast versus cosmological redshift for the system of Eqs. \eqref{ddotdelta44}- \eqref{ddotv44} for $n >1$ for long-wavelength ($K= 0.01$) for set II.}
    \label{Figure5}
\end{minipage} 
\qquad
 \begin{minipage}{0.5\textwidth}
\includegraphics[width=0.9\textwidth]{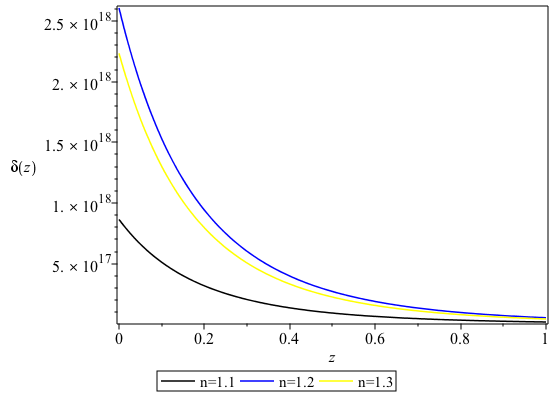}
   \caption{The growth of the  density contrast versus cosmological redshift for the system of Eqs. \eqref{ddotdelta44}- \eqref{ddotv44} for $n >1$  for short-wavelength ($K= 10^5$) for set II.}
    \label{Figure6}
\end{minipage} 
\qquad
 \begin{minipage}{0.5\textwidth}
\includegraphics[width=0.9\textwidth]{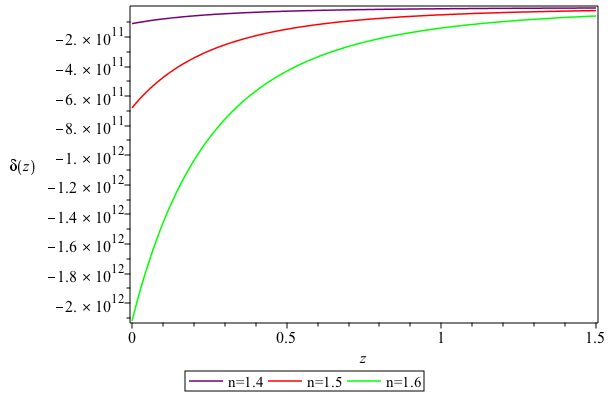}
   \caption{The growth of the density contrast versus cosmological redshift for the system of Eqs. \eqref{ddotdelta44}- \eqref{ddotv44} for $n >.31$  for long-wavelength ($K= 0.001$)   for set II.}
    \label{Figure7}
\end{minipage} 
 \qquad
 \begin{minipage}{0.5\textwidth}
\includegraphics[width=0.9\textwidth]{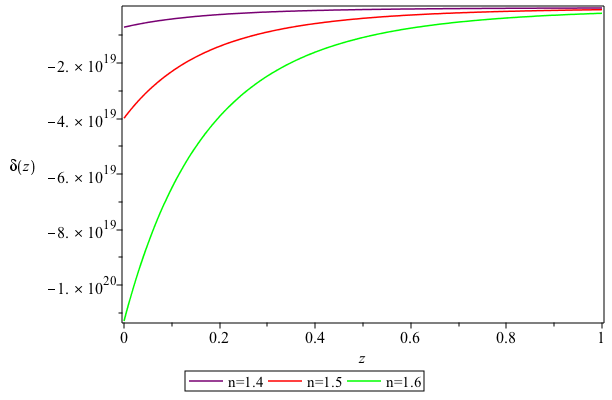}
   \caption{The growth of the  density contrast versus cosmological redshift for the system of Eqs. \eqref{ddotdelta44}- \eqref{ddotv44} for $n >1.3$ for short-wavelength ($K= 10^5$) for set II.}
    \label{Figure8}
\end{minipage} 
\end{figure}

\begin{figure}[H]
\begin{minipage}{0.5\textwidth}
\includegraphics[width=0.9\textwidth]{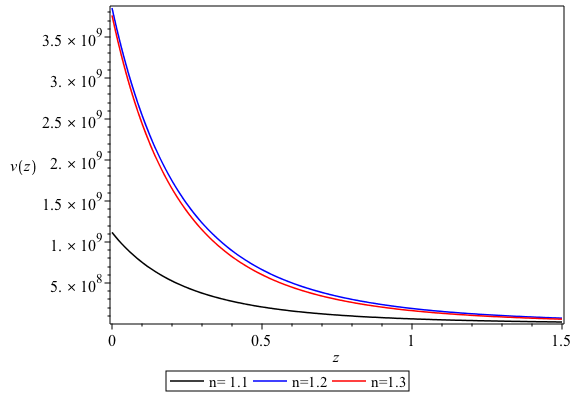}
    \caption{The growth of the  velocity contrast versus cosmological redshift for the system of Eqs. \eqref{ddotdelta44}- \eqref{ddotv44} for $n >1$  for set II.}
    \label{Figure9}
\end{minipage}
\qquad
\begin{minipage}{0.47\textwidth}
\includegraphics[width=0.9\textwidth]{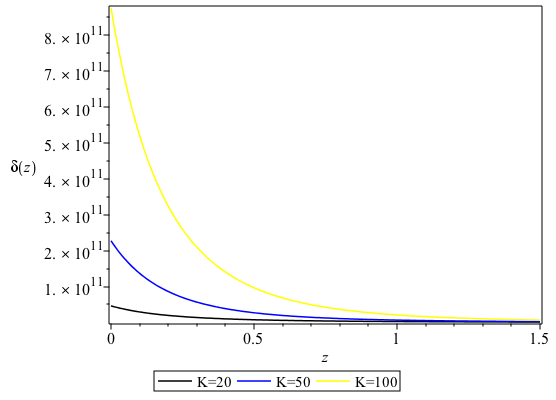}
    \caption{The growth of the density  contrast versus cosmological redshift for the system of Eqs. \eqref{ddotdelta44}- \eqref{ddotv44} for $n =1.1$  and different values of $K$  for set II.}
    \label{Figure11}
\end{minipage}
\end{figure}

The numerical results Eqs. \eqref{dotz12}- \eqref{dotv1112} for the short- and long- wavelength modes are presented in Figs. \ref{fig1f} - \ref{Figure3f}. We have tried to study the behaviour of the velocity and the  density contrast for different  $n$ ranges, i.e., for $n<1 $, $n$ closes to GR and $n>1$. The velocity and the density contrast are decaying for values of $n<1$ and for $n \geq 1.4$. We only noticed the growth of the velocity and the density fluctuations  for values of $1\leq n\leq 1.3$ as presented in Figs. \ref{Figure1f} - \ref{Figure3f}. The growth of the velocity and the density perturbation is highly nonlinear compared to the GR results.
The numerical results Eqs. \eqref{ddotdelta44} and \eqref{ddotv44} for the short- and long- wavelength modes for different sets of the initial conditions are presented in the following figures.   For instant, the numerical results for set I of the initial conditions as presented in Figs. \ref{fig1} - \ref{Figure4}. We have noticed that the velocity and  density contrast are decaying  for values of $0.5 < n\leq 0.99 $ and for the case of $n = 1$, the numerical results of GR are recovered as in Figs. \ref{fig1} - \ref{fig2}.  For values of $ n>1$ as they are presented in Figs. \ref{Figure1} - \ref{Figure2}, we can see the growth of the density contrast for both short- and  long-wavelengths and the only difference  appears in the amplitudes of the density contrast,  in Fig. \ref{Figure1}  we can see the growth of the density contrast for long-wavelength with less amplitudes compared to the short-wavelength results as presented in Fig. \ref{Figure2}. We also have to mention that the  growth of the velocity contrast does not depend on the wave number $K$. We also presented the behaviour of the matter density contrast for $n =1.1$ and different values of the wave number $K$ as presented in Fig. \ref{Figure4}, and we noticed the growth of the density contrast with increasing the values of $K$. We present the behaviour of the density fluctuations in Table. \ref{Table1}
The numerical results for set II of the initial conditions are presented in  Figs. \ref{Figure5} - \ref{Figure11}.  We noticed that for values of $0.5 < n\leq 0.99$ the velocity and  density contrast are decaying, and the GR are recovered for $n=1$ as in Figs.  \ref{fig:GRV} - \ref{fig:GRd}. We  also noticed  the growth of the velocity and the density contrast for $ 1< n\leq 1.3$  as presented in Figs. \ref{Figure5}, \ref{Figure6} and \ref{Figure9}. For  $ 1.4\leq n $, we observe unrealistic behaviour of the density contrast  for both short- and long- wavelengths  as presented in Figs. \ref{Figure7} - \ref{Figure8}. We also noticed that  the velocity grows at $n=1.6$, and start to decay again for values of $n>1.6$. We present the behaviour of the density fluctuations in Table. \ref{Table2}
\begin{table}[H]
\caption{The behavior of $\delta(z)$ for set I.}
\begin{tabular}{lllll}
Range of $n$ & Density contarst &  \\
 \hline
 $0.5 < n \leq 0.99 $ & decreasing & &  \\
 \hline
 $n\geq 1 $ &  increasing &  \\
  \hline
\end{tabular}
\label{Table1}
\end{table}
\begin{table}[H]
\caption{The behavior of $\delta(z)$ for set II.}
\begin{tabular}{lllll}
Range of $n$ & Density contarst &  \\
 \hline
 $0.5 < n \leq 0.99 $ & decreasing & &  \\
 \hline
 $1 \leq n\leq 1.3 $ &  increasing &  \\
  \hline
  $n \geq 1.4$ &  decreasing &  \\
  \hline
  \end{tabular}
\label{Table2}
\end{table}
\subsubsection{Case II: Quasi-static approximations}
\begin{figure}[H]
\begin{minipage}{0.5\textwidth}
\includegraphics[width=0.9\textwidth]{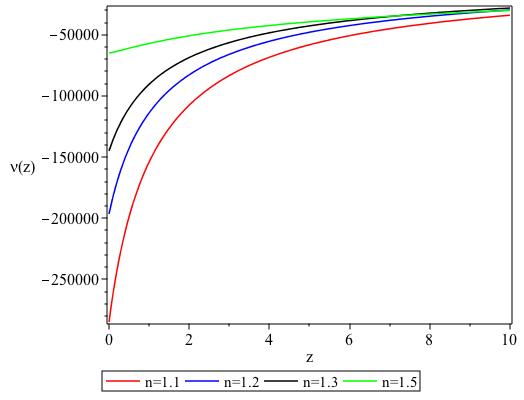}
    \caption{The growth of the velocity contrast versus cosmological redshift for Eq. \ref{v121}
    for $ 1\leq n\leq 1.5$ for set I.}
    \label{fig11q}
\end{minipage}
 \qquad
\begin{minipage}{0.5\textwidth}
\includegraphics[width=0.9\textwidth]{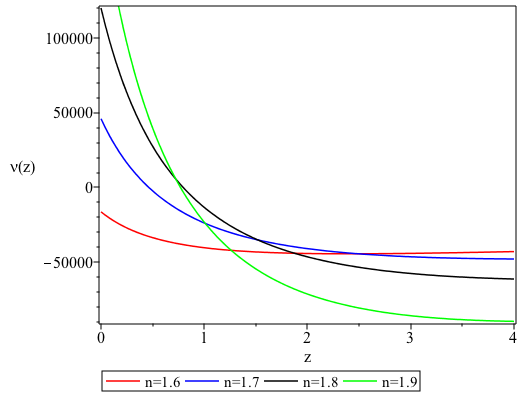}
    \caption{The growth of the velocity contrast versus cosmological redshift for Eq. \eqref{v121}
    for $ 1.6 \leq n$ for set I.}
   \label{fig12q}
    \end{minipage}
    \qquad
\begin{minipage}{0.5\textwidth}
\includegraphics[width=0.9\textwidth]{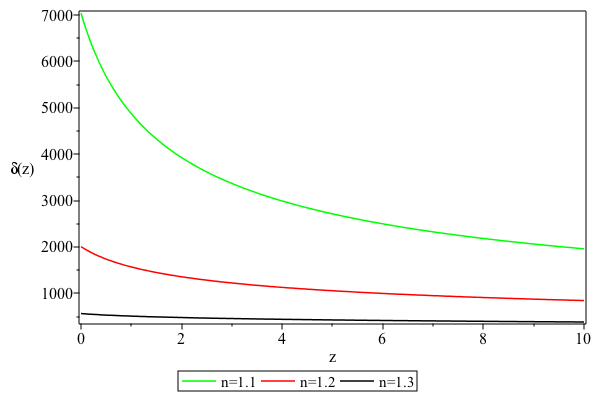}
    \caption{The growth of  the density contrast versus cosmological redshift for Eq. \eqref{D121}
    for $ 1\leq n\leq 1.3$ for set I.}
    \label{fig121q}
\end{minipage}
\qquad
\begin{minipage}{0.5\textwidth}
\includegraphics[width=0.9\textwidth]{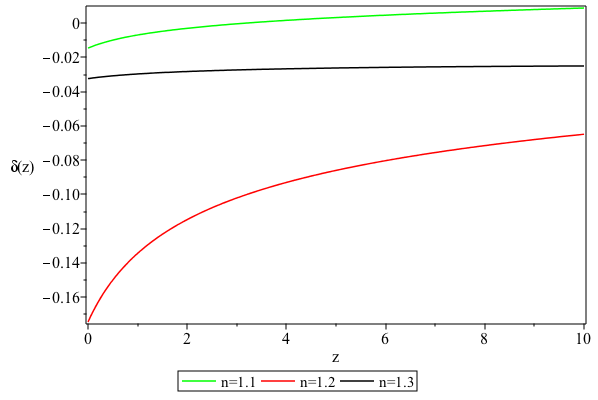}
    \caption{The growth of density contrast versus cosmological redshift for Eq. \eqref{D121}
    for $ 1\leq n\leq 1.3$ for set II.}
    \label{fig111q}
\end{minipage}
 \qquad
\begin{minipage}{0.5\textwidth}
\includegraphics[width=0.9\textwidth]{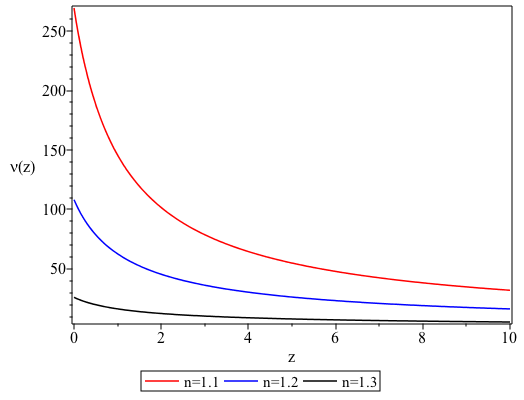}
    \caption{The growth of the velocity contrast versus cosmological redshift for Eq. \eqref{v121}
    for $ 1 \leq n\leq 1.3 $ for set II.}
   \label{fig14q}
    \end{minipage}
    \qquad
   \begin{minipage}{0.5\textwidth}
\includegraphics[width=0.9\textwidth]{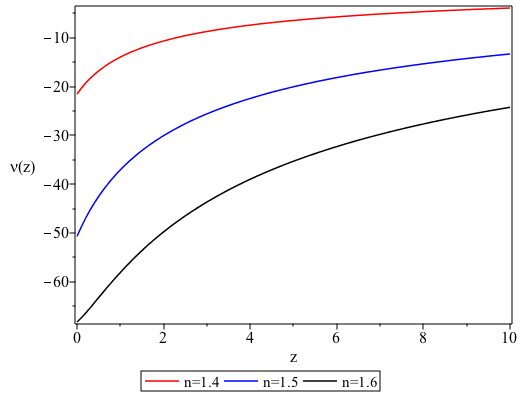}
    \caption{The growth of the velocity contrast versus cosmological redshift for Eq. \eqref{v121}
    for $ n \geq 1.4 $ for set II.}
   \label{fig14q}
    \end{minipage} 
\end{figure}

In this sub-subsection, we present the results of the velocity and density  contrast in the quasi-static approximations Eqs.  \eqref{v12}-\eqref{D12}
 We apply the same technique as GR limits and we find first the exact solution of velocity contrast and then the density contrast. So, the exact solution of Eq. \eqref{v12} is given as\\
\begin{eqnarray}\label{v121}
 &&V(z)= {\it c_5}\, \left( 1+z \right) ^{\alpha_{+}}+{\it c_6}\, \left( 1+z \right) 
^{{\alpha_{-}}}\;, \label{quasi6}
\end{eqnarray}
where
 \begin{eqnarray} 
&& \alpha_{\pm} = \frac {\mathcal{Y} \eta_{1}}{4} + \frac{1}{4}\pm \frac{ \sqrt { \eta^{2}_{1} \mathcal{Y}^{2}  -24 \eta_{2} \mathcal{Y}\tilde{\Omega}_{m} +2 \eta_{1} \mathcal{Y} +24  \tilde{\Omega}_{m} +1}} {4}  \;, \\
 &&\eta_{1}=  \frac{2(20n-15) (3-7n)}{3\Big( 32n^{2} -41n +15\Big)}\;, \hspace{.3cm}
 \eta_{2}=  \frac{n(20n-15)}{3\Big( 32n^{2} -41n +15\Big)}\;.
 \end{eqnarray}
  After we computed the integration constants $c_5$ and $c_6$ by imposing the initial conditions, the exact solution for density contrast Eq. \eqref{D12} for $K=0$ is given as
  \begin{eqnarray}\label{D121}
&\Delta_m(z) = {\it c_7}\left( 1+z \right) ^{\alpha_{1}} +  {\it c_8}\left( 1+z \right) ^{\alpha_{2}} \nonumber \\&- 
 \frac{ 2n^{3}}{3\Big( -\frac{9}{2} +  n^{2}\big( \mathcal{Y} \alpha_{+} -2\alpha^{2}_{+} + 3\tilde{\Omega}_{m} + 3\alpha_{+}\big) + n\big( \frac{3\mathcal{Y}}{2} -6\alpha_{+} +\frac{9}{2}\big)\Big)\Big( -\frac{9}{2} +  n^{2}\big( \mathcal{Y} \alpha_{-} -2\alpha^{2}_{-} + 3\tilde{\Omega}_{m} + 3\alpha_{-}\big) + n\big( \frac{3\mathcal{Y}}{2} -6\alpha_{-} +\frac{9}{2}\big)\Big)}\times   \nonumber\\ & \Big\lbrace \Big( \mathcal{Y} \eta_{3} ( \tilde{\Omega}_{m} \eta_{4} -\eta_{5} \alpha_{+})+ 2\alpha_{+} + \eta_{6} + 9\tilde{\Omega}_{m} \Big)  \Big(\Big( -\frac{9}{2} +  n^{2}\big( \mathcal{Y} \alpha_{-} -2\alpha^{2}_{-} + 3\tilde{\Omega}_{m} + 3\alpha_{-}\big) \nonumber \\&+ n\big( \frac{3\mathcal{Y}}{2} -6\alpha_{-} +\frac{9}{2}\big)\Big) {\it c_5} (1+z)^{\frac{(2\alpha_{+} n+3)}{2n}} \Big) \nonumber \\& + \Big( \mathcal{Y} \eta_{3} ( \tilde{\Omega}_{m} \eta_{4} -\eta_{5} \alpha_{-})+ 2\alpha_{-} + \eta_{6} + 9\tilde{\Omega}_{m} \Big)  \Big(\Big( -\frac{9}{2} +  n^{2}\big( \mathcal{Y} \alpha_{+} -2\alpha^{2}_{+} + 3\tilde{\Omega}_{m} + 3\alpha_{+}\big) \nonumber \\&+ n\big( \frac{3\mathcal{Y}}{2} -6\alpha_{+} +\frac{9}{2}\big)\Big) {\it c_6} (1+z)^{\frac{(2\alpha_{-} n+3)}{2n}} \Big)\Big\rbrace\;,
 \end{eqnarray}
where 
\begin{eqnarray}
&&\alpha_{1,2}= \frac{\mathcal{Y}}{4} + \frac{3}{4} \pm \frac{\sqrt{\mathcal{Y}^{2} + 24 \tilde{\Omega}_{m} + 6 \mathcal{Y} +9}}{4}\\&&
\eta_{3}=  \Big( 6\tilde{\Omega}_{m} + 6\mathcal{X}  +\frac{(96n-27-31n^{2})}{n^{2}}\Big) \;, \hspace{.3cm}
\eta_{4}= \frac{2n^{2}}{(32n^{2} -41n +15)}\;,  \\&&
\eta_{5}=  \frac{2n(3-7n)}{3(32n^{2} -41n +15)}\;, \hspace{.3cm}
\eta_{6}= \Big( 3+ 3\tilde{\Omega}_{m} -6\mathcal{X} +\frac{27(n-1) (2n-1)}{2n^{2}}\Big)\;.
\end{eqnarray}
 In the following figures we present the results of the velocity and the density contrast in the quasi-static approximations Eqs.  \eqref{v121} and \eqref{D121} for different $n$ ranges. For instant, for set I of the initial conditions, the GR results are recovered as presented in Figs. \ref{fig1} - \ref{fig2}.  The velocity is decaying and  for values of $ 1 \leq n \leq 1.5$ as they are presented in Fig. \ref{fig11q}, and we see the growth of the velocity contrast only for values of $1.6 \leq n \leq 1.9$ as presented in Fig. \ref{fig12q}. We could only evaluate the results for the density contrast only for long-wavelength ($K=0$) , we see the growth of the density contrast only at values of $ 1 \leq n \leq 1.3$ as in Fig. \ref{fig121q}. We also presented the results for set II of the initial conditions in Figs. \ref{fig111q} - \ref{fig14q}, and we noticed the growth of the velocity contrast and the decaying of the density contrasrt for $ 1 \leq n \leq 1.3$ . By comparing these results with the results obtained by solving the full system at the same choice of initial conditions (sets I and II), we can conclude that the quasi-static approximation does not seem to be applicable in the quasi-Newonian framework.
\section{Conclusions}\label{sec8}
This work presented a detailed analysis of scalar cosmological perturbations in $f(R)$ as a scalar-tensor theory of gravity theory using the $1 + 3$ covariant gauge-invariant approach. We explored the integrability conditions of the so-called quasi-Newtonian cosmological models in the context of $f(R)$ as scaler-tensor theory of gravity. We showed that for such cosmological models to exist, they must satisfy certain integrability conditions on the generalized Einstein field equations. The two integrability conditions derived and presented here allow us to describe a consistent evolution of the linearised field equations of quasi-Newtonian universes. We defined the scalar gradient variables and derived the corresponding evolution equations. We derived the complete set of the first- and the second-order evolution equations of these perturbations. The harmonic decomposition approach is applied to
these equations in order to solve this complicated system of differential equations. After getting a complete set of the perturbation equations, We studied the behaviour of matter energy density perturbations with redshift for different ranges of $n$ by considering  $R^{n}$ models.  We introduced the so-called quasi-static approximation to study the approximated solutions on small scales. We have shown the ranges of $n$ for which the perturbation amplitudes $\delta(z)$  grow or decay. For instance, the numerical solution  of the first-order perturbation equations  shows the growth of the density contrast for values of $1\leq n \leq 1.3$ $ v _{in}= v(z_{in}\simeq 1100)= 10^{-5}$,  $ \Phi_{in}= \Phi(z_{in}\simeq 1100)= 10^-5$ , $\xi _{in}= \xi(z_{in}\simeq 1100)=10^{-5}$, $\mathcal{A} _{in}= \mathcal{A}(z_{in}\simeq 1100)= 10^{-5}$,  $\mathcal{Z} _{in}=\mathcal{ Z}(z_{in}\simeq 1100)= 10^{-5}$  and $\Delta _{in}= \Delta _{m}(z_{in}\simeq 1100)= 10^{-5}$. We also found the numerical solution of the whole system of second-order equations and we escalated our analysis for different sets of initial conditions for the short- and long- wavelength modes. For instant,  for set I as presented in Figs. \ref{fig1} - \ref{Figure4}. We have noticed that the velocity and  density contrast are decaying  for values of $0.5 < n\leq 0.99 $ and for the case of $n = 1$, the numerical results of GR are recovered as in Figs. \ref{fig1} - \ref{fig2}.  For values of $ n>1$ as they are presented in Figs. \ref{Figure1} - \ref{Figure2}, we can see the growth of the density contrast for both short- and  long-wavelengths. We also presented the behaviour of the matter density contrast for $n =1.1$ and different values of the wave number $K$ as presented in Fig.\ref{Figure4}, and we noticed the growth of the density contrast with increasing the values of $K$.  for set II   We  noticed  the growth of the velocity and the density contrast for $ 1< n\leq 1.3$  as presented in Figs. \ref{Figure5}, \ref{Figure6} and \ref{Figure9}. For  $ 1.4\leq n $, we observe unrealistic behaviour of the density contrast  for both short- and long- wavelengths  as presented in Figs. \ref{Figure7} - \ref{Figure8}.  While On-the other hand, based on the quasi-static approximation, we could only evaluate the exact solutions for the density contrast at $K=0$. For instant, for set I of the initial conditions, we see the growth of the density contrast only at values of $ 1 \leq n \leq 1.3$ as in Fig. \ref{fig121q}, while we noticed a decaying behaviour at the same values of $n$ or set II of the initial conditions as in Fig. \ref{fig14q}. By comparing these results with the results obtained by solving the full system at the same choice of initial conditions (sets I and II), we can conclude that the quasi-static approximation does not seem to be applicable here.
\section{Acknowledgments}
HS gratefully acknowledges the financial support from the Mwalimu Nyerere African Union scholarship and the National Research Foundation (NRF) free-standing scholarship. AA acknowledges that this work is based on the research supported in part by the NRF of South Africa.
\newpage
\appendix
\section*{Appendix}
\renewcommand{\theequation}{\thesection.\arabic{equation}}
\renewcommand{\thesection}{A} 
Some of the following linearised identities which hold for all scalars $f$, vectors $V_a$ and tensors $S_{ab}=S_{\langle ab\rangle}$, have been used in this paper:
\begin{eqnarray}\label{a1}
&&\eta^{abc}\tilde{\nabla}_{b}\tilde{\nabla}_{c}f=0\; ,\\
&&\label{a2}
\left( \tilde{\nabla}_{\langle a}\tilde{\nabla}_{b\rangle}f\right)^{\cdot}=\tilde{\nabla}_{\langle a}\tilde{\nabla}_{b\rangle} \dot{f} -\dfrac{2}{3} \theta \tilde{\nabla}_{\langle a}\tilde{\nabla}_{b\rangle}f +\dot{f}\tilde{\nabla}_{\langle a}A_{b\rangle}\; ,\\
&&\label{a3}
\left( \tilde{\nabla}_{a}f\right)^{\cdot}= \tilde{\nabla}_{a}\dot{f}-\dfrac{1}{3}\theta\tilde{\nabla}_{a}f+ \dot{f}A_{a}\; ,\\
&&\label{a4}
\tilde{\nabla}^{b}\tilde{\nabla}_{<a}A_{b>}= \dfrac{1}{2}\tilde{\nabla}^{2}A_{a}+\dfrac{1}{6}\tilde{\nabla}_{a}\tilde{\nabla}^{c}A_{c}+ \dfrac{1}{3}(\rho- \dfrac{1}{3} {\theta}^{2})A_{a}\;,\\
&&\label{a5}
\left(\tilde{\nabla}^{2}f\right)^{\cdot}=\tilde{\nabla}^{2}\dot{f}-\dfrac{2}{3}\theta \tilde{\nabla}^{2}f+ \dot{f}\tilde{\nabla}^{a}A_{a}\; ,\\
&&\label{a6}
\tilde{\nabla}_{[a}\tilde{\nabla}_{b]}v_{c}=\dfrac{1}{3}\left(\dfrac{1}{3}\theta^{2}-\rho\right)v_{[a}h_{b]c}\; ,\\
&&\label{a7}
\tilde{\nabla}_{[a}\tilde{\nabla}_{b]}S^{cd}=\dfrac{2}{3}\left(\dfrac{1}{3}\theta^{2}-\rho\right)S_{[a}^{(c}h_{b]}^{d)}\; ,\\
&&\label{a8}
\tilde{\nabla}^{a}\left( \eta_{abc}\tilde{\nabla}^{b}v^{c}\right)=0\; ,\\
&&\label{a9}
\tilde{\nabla}_{b}\left( \eta^{cd\langle a}\tilde{\nabla}_{c}S^{b\rangle}_{d}\right)= \dfrac{1}{2}\eta^{abc}\tilde{\nabla}_{b}\left(\tilde{\nabla}_{d}S^{d}_{c}\right)\; ,\\
&&\label{a10}
\tilde{\nabla}^{2}(\tilde{\nabla}_{a}f)= \tilde{\nabla}_{a}(\tilde{\nabla}^{2}f)+\dfrac{2}{3}(\rho-\dfrac{1}{3}\theta^{2})\tilde{\nabla}_{a}f+ 2\dot{f}\eta_{abc}\tilde{\nabla}^{b} \omega^{c}\;.
\end{eqnarray}
\bibliographystyle{iopart-num}
\bibliography{./references}

\end{document}